\documentclass[amsmath,showpacs,aps,prb,twocolumn,superscriptaddress,floatfix]{revtex4-1}
\usepackage{amsmath}
\usepackage{dcolumn}
\usepackage{amsfonts}
\usepackage{amssymb}
\usepackage{graphicx}
\usepackage{natbib}
\usepackage{color}
\newcommand{\bs}{\boldsymbol}
\usepackage{subfigure}

\usepackage[colorlinks=true,citecolor=blue,linkcolor=blue]{hyperref}

\begin{document}
\title{Lattice distortion effects on topological phases in (LaNiO$_3$)$_2$/(LaAlO$_3$)$_N$ heterostructures grown along the [111] direction}
\author{Andreas R\"uegg} 
\affiliation{Department of Physics, University of California, Berkeley, CA 94720, USA} 
\author{Chandrima Mitra}
\affiliation{Department of Physics, The University of Texas at Austin, Austin, Texas 78712, USA}
\author{Alexander A.~Demkov}
\affiliation{Department of Physics, The University of Texas at Austin, Austin, Texas 78712, USA}
\author{Gregory A. Fiete}
\affiliation{Department of Physics, The University of Texas at Austin, Austin, Texas 78712, USA}
\date{\today}
\begin{abstract}
We theoretically investigate the influence of lattice distortion effects on possible topological phases in (LaNiO$_3$)$_2$/(LaAlO$_3$)$_N$ heterostructures grown along the [111] direction.  At the Hartree-Fock level, topological phases originate from an interaction-generated effective spin-orbit coupling that opens a gap in the band structure. For the unstrained system, there is a quadratic band touching at the $\Gamma$ point at the Fermi energy for spin unpolarized electrons and Dirac points at K, K$'$ at the Fermi energy for fully spin polarized electrons. Using density functional theory we first show that the quadratic band touching and Dirac points are remarkably stable to internal strain-induced out-of-plane distortions and rotations of the oxygen octahedra, which we compute. The lack of a gap opening implies that the mean-field predictions for topological phases for both the polarized and unpolarized systems are little affected by internal strain and lattice relaxations. On the other hand, we also discuss two types of lattice distortions which have an important effect on the electronic structure. First, an external strain imposed along the [001] cubic direction can open a gap at the $\Gamma$ point, thereby stabilizing a trivial insulating phase in the spin unpolarized system. However, it leaves the Dirac points intact. As a result, the Hartree-Fock calculation for an effective model using parameters relevant to LaNiO$_3$ finds that symmetry-breaking strain favors a phase with polarized orbitals and antiferromagnetic spin order, but leaves earlier predictions for a zero-magnetic field topological quantum Hall state essentially unchanged. Second, we identify a possible breathing distortion of the oxygen cages stabilized by correlation effects. Such a distortion opens a gap at the Dirac points and we demonstrate that it would compete with the topological phase in the fully spin polarized system. Taken together, our results suggest that distortion effects in thin films grown along the [111] direction in perovskites have rather different consequences as compared to those grown along [001].

\end{abstract}


\maketitle

\section{Introduction}

The experimental discovery of two\cite{Konig:sci07,Roth:sci09,Chang:sci13} and three\cite{Hsieh:nat08,Hsieh:sci09,Xia:np09} dimensional topological insulators has intensified the study of time-reversal preserving and time-reversal breaking topological phases.\cite{Hasan:rmp10,Moore:nat10,Qi:rmp11} From a practical standpoint it is important to identify new materials with topological properties\cite{Zhang:np09,Chadov:nm10,JWang:prl11,Feng:prl11,Zhang:prl11,Xiao:prl10,Lin:prl10,Chen:prl10,Yan:epl10,Yan:prb10}
and better understand the role that interactions may play in driving ``conventional" and ``exotic" topological phases if the full potential of these materials is to be realized in applications.\cite{Young:prb08,Ran:prl08,Raghu:prl08,Zhang:prb09,Wen:prb10,Li:np10,Swingle:prb11,Maciejko:prl10,Levin:prb12,Levin:prl09,Moore:prl08,Neupert:prb11,Witczak-Krempa:prb10,Liu:prb10}

Because of the intrinsic role that interactions play in the physics of transition metal oxides they have emerged as an important frontier in topological insulator research.\cite{Shitade:prl09,Pesin:np10,Yang_Kim:prb10,Kargarian:prb11,Wan:prb11,Ruegg:prl12,Kargarian:prb12,Reuther:prb12,Go:prl12,Yang:prb11b,Witczak:prb12,Kargarian:prl13}  An especially promising area in the search for topological phases is the interface of correlated oxides\cite{Xiao:nc11,Ruegg11_2,Ruegg:prb12,Yang:prb11a,Wang:prb11,Hu:prb12}--in part due to the large degree of ``tunability" in such systems, but also because the interface is a natural physical structure in devices.  Independent of the interest in topological phases, oxide interfaces have proven to be an intrinsically rich system for realizing correlated phases.\cite{Hwang:nm12,Mannhart:sci10,Mannhart:mrb08,Zubko:arcmp11,Boris:sci11,Benckiser:nm11,Chakhalian:prl11,Lui:prb11,Son:apl10,Son_2:apl10} 

Because of their relative abundance (and therefore relatively low-cost) and wide range of electronic phases, transition metal oxides with the perovskite structure ABO$_3$, where A is usually a rare earth element, B is a transition metal, and O is oxygen have undergone intensive study.  The undistorted perovskite has a relatively simple cubic structure with natural cleave planes along the [001] and equivalent directions, which makes it a natural direction for growth.  However, experiments on thin films grown along the [001] direction show significant and sometimes anisotropic (with respect to ``compression" and ``stretching") lattice strain effects on the electronic properties.\cite{Hwang:nm12,Mannhart:sci10,Mannhart:mrb08,Zubko:arcmp11,Boris:sci11,Benckiser:nm11,Chakhalian:prl11,Lui:prb11,Son:apl10,Son_2:apl10,Hwang:2013}  While such strain-induced electronic effects may turn out to be important for some applications,\cite{Chakhalian:nm12,Rondinelli:am12} it is crucial to pursue alternate material growth routes, such as interface/thin film growth along the [111] direction, to better understand the relation between strain and electronic structure, and to identify cases where strain effects appear to be minimal.\cite{Oja:prb08,Berger:prl11,Tagantsev:prb01}
 
Already, a number of theoretical studies suggest that thin films, particularly bilayers and trilayers, grown along the [111] direction under conditions of minimal strain are promising for realizing topological phases.\cite{Xiao:nc11,Ruegg11_2,Ruegg:prb12,Yang:prb11a,Wang:prb11,Hu:prb12,Lado:2013} The first experimental searches in (LaNiO$_3$)$_2$/(LaAlO$_3$)$_N$ have been undertaken,\cite{Middey:apl12} but so far have provided inconclusive results for the presence of topological phases.  In addition to the topological phases addressed here, other theoretical studies have considered the [111] interface/thin films in the limit of strong electron correlations where a local moment model is a natural starting point.\cite{Okamoto:prl13,Dong:2013} Further experimental studies on different materials have shown growth in the [111] direction in perovskites is achievable in spite of its highly polar nature.\cite{Herranz:sp2012,Gibert:nm2012,Blok:apl11}

Of particular importance for topological phases in the (LaNiO$_3$)$_2$/(LaAlO$_3$)$_N$ heterostructures grown along [111] is the electronic structure of the $e_g$ bands (the nominal charge of the Ni is 3+, which leads to a $t_{2g}^6e_g^1$ electronic configuration). Here, the growth direction enters in an important way: the two Ni ions of the LaNiO$_3$-bilayer form a buckled honeycomb lattice which gives rise to features in the band structure  known from graphene and other hexagonal systems.\cite{Xiao:nc11} In particular, in the simplest tight-binding model for an undistorted lattice, there is a flat-band touching a (locally) quadratically dispersing band at $\Gamma$ at an energy corresponding to the Fermi energy of spin unpolarized electrons at quarter filling.\cite{Xiao:nc11,Ruegg11_2,Yang:prb11a} The flat-band touching point is robust within density functional theory (DFT) with the local density approximation (LDA) for an undistorted lattice,\cite{Ruegg:prb12} and plays a key role in abetting interaction-driven topological phases at weak coupling.\cite{Sun:prl09,Wen:prb10,FZhang:prl11,Ruegg11_2,Yang:prb11a} On the other hand, the electronic structure also features linear touching points (Dirac points) located at the corners  K and K$'$ of the hexagonal Brillouin zone. They are at an energy relevant for the fully spin polarized system at quarter filling. Such a ferromagnetic state is indeed found within LSDA+$U$ and Hartree-Fock for certain values of interaction parameters.\cite{Yang:prb11a,Ruegg:prb12} Though the Dirac points are not perturbatively unstable to interactions in two dimensions, topological phases can result from a gap opening beyond a critical interaction strength.\cite{Sun:prl09,Wen:prb10,Raghu:prl08}

\begin{figure}[htb]
\centering
\subfigure[\ Supercell]{
\label{fig1a}
\includegraphics[width=0.99\linewidth]{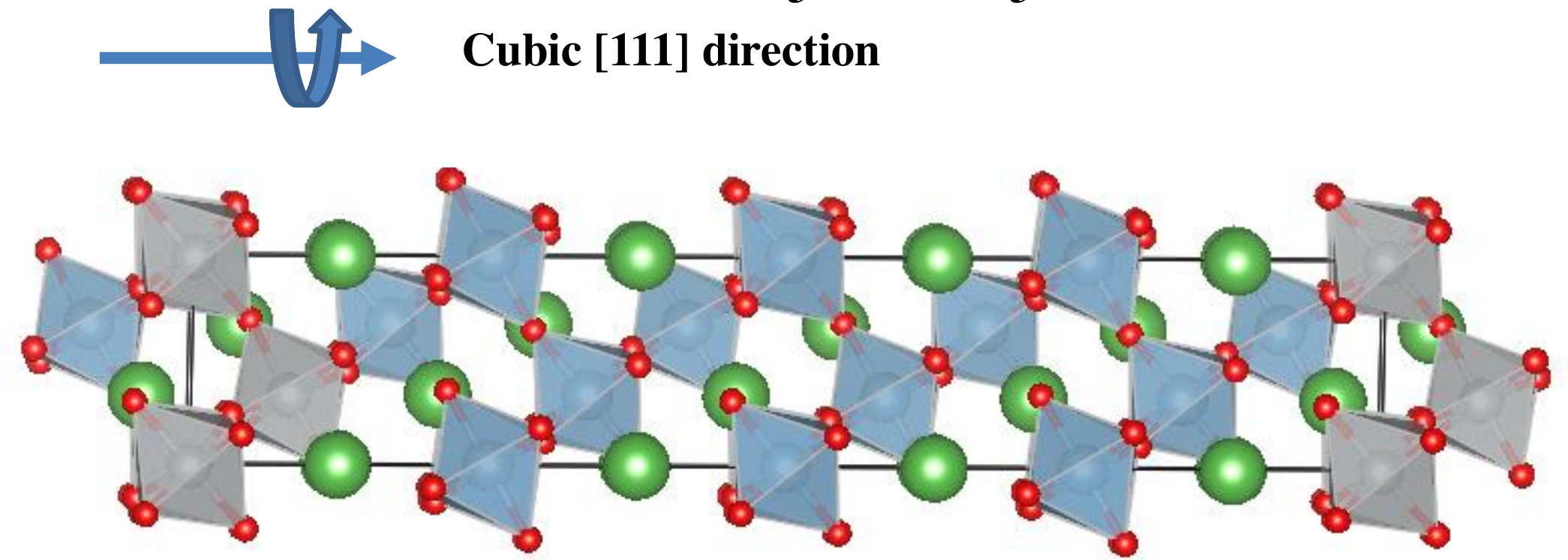}}
\subfigure[\ Octahedral tilts and rhombohedral cell angle]{
\label{fig1b}
\includegraphics[width=0.5\linewidth]{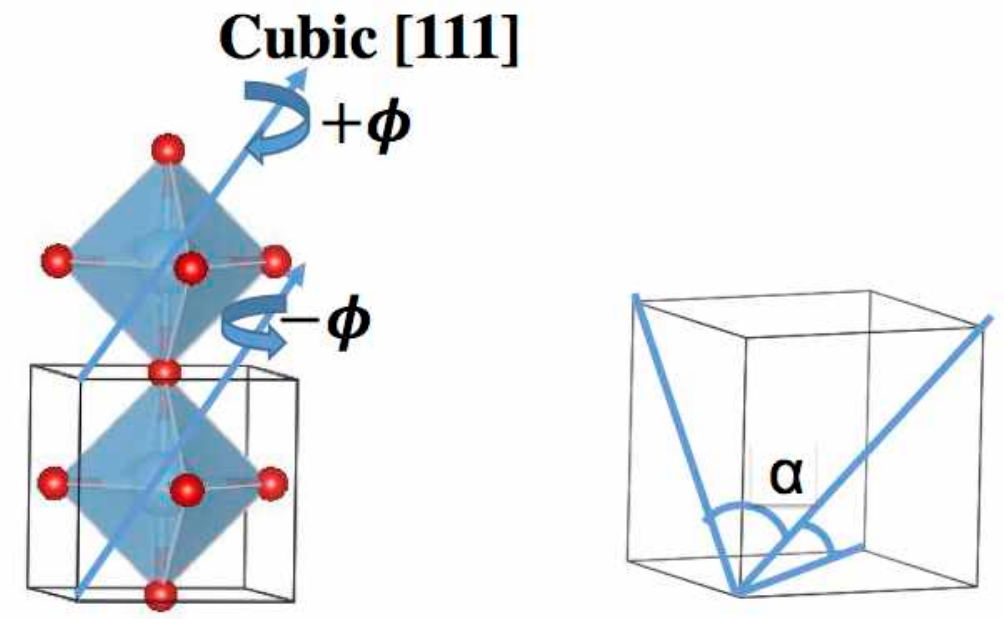}}
\hfill
\subfigure[\ View along {[111]}]{
\label{fig1c}
\includegraphics[width=0.45\linewidth]{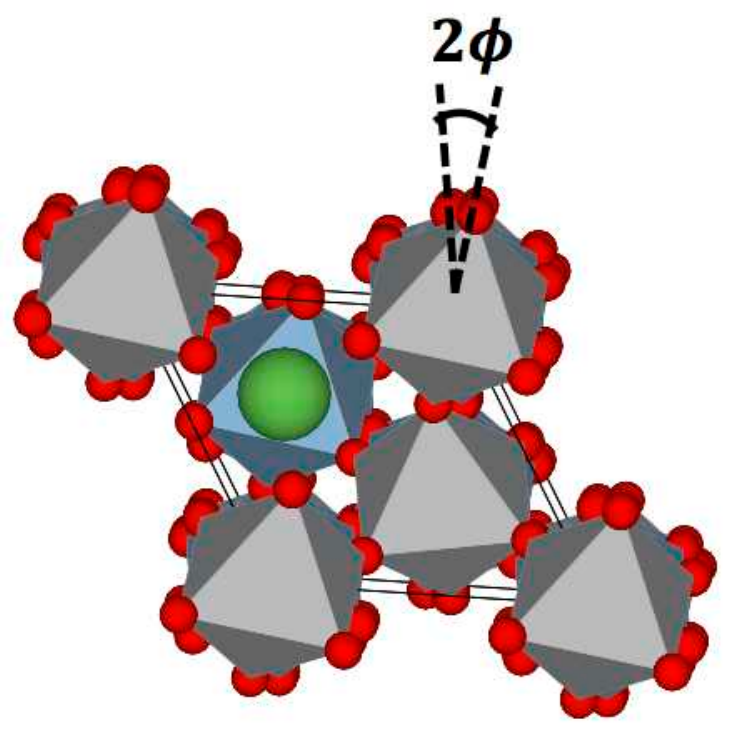}}
\caption{(Color online.) (a) Supercell used in the density functional theory calculations reported in this work. Octahedral rotations are shown. Red spheres represent oxygen atoms, and green spheres represent La atoms. Nickel atoms sit at the centers of the octahedral oxygen cages. (b) The pattern of the octahedral tilts present in the fully relaxed structure and a sketch of the rhombohedral cell angle $\alpha$. (c) A view along the [111] direction showing the octahedral rotations. Because the octahedral rotations preserve the trigonal point group symmetry, the quadratic band touching and Dirac points are preserved. This, in turn, implies the predictions for interaction-driven topological phases in the (LaNiO$_3$)$_2$/(LaAlO$_3$)$_N$ system remain qualitatively unchanged. As we discuss, a symmetry-breaking external strain is required for a qualitative change.}
\label{fig:}
\end{figure}

In this paper, we investigate how robust the electronic structure of the (LaNiO$_3$)$_2$/(LaAlO$_3$)$_N$ system is against various lattice distortions and how potential topological phases are affected by such perturbations. We carry out this study using DFT with LDA and GGA (generalized gradient approximation), which are believed to provide reasonably accurate results for the band features of metallic LaNiO$_3$. Electron-electron interaction effects are discussed both within LSDA+$U$ and within the Hartree-Fock approximation using the LDA band structure as input. While Hartree-Fock theory is not expected to provide a quantitatively accurate description of LaNiO$_3$, our focus here is on the specific issue of whether the previously identified instabilities towards an insulating topological phase in the presence of interaction\cite{Ruegg11_2,Ruegg:prb12,Yang:prb11a} persists under DFT-{\em computed} band structure with internal and external lattice strain in the thin-film geometry. On this question, Hartree-Fock theory should be more reliable; in model Hamiltonian systems, functional renormalization group studies (a complimentary ``unbiased" approach) support the Hartree-Fock mean-field predictions of interaction-generated topological phases.\cite{Raghu:prl08,Uebelacker:2011} 

One of our main results is that the quadratic band touching at the $\Gamma$ point and the Dirac points at K, K$'$ are remarkably stable to internal strain in [111] grown films. To show this we compute the fully-relaxed lattice structure, corresponding oxygen tilt angles, and resulting band structure.  By fitting to a tight-binding model, we see that hopping parameters are uniformly reduced by roughly 10-15\%, but there is little other change to the band structure. This small change to the kinetic energy results in a very small numerical change to the previously obtained Hartree-Fock phase diagrams.\cite{Ruegg:prb12} The central conclusion of our study of internal strain on the (LaNiO$_3$)$_2$/(LaAlO$_3$)$_N$ system is that it has negligible effects on the results previously obtained for the unstrained system.\cite{Ruegg11_2,Ruegg:prb12,Yang:prb11a} In particular, earlier proposals\cite{Ruegg11_2,Ruegg:prb12,Yang:prb11a} for topological phases in the (LaNiO$_3$)$_2$/(LaAlO$_3$)$_N$ system are essentially unaffected.

On the other hand, we also identify lattice distortions which {\em do} have an important effect on the electronic structure. First, we show it is possible to forcefully ``push" the system to open a gap in the band structure from an externally applied strain along the [001] cubic direction. We find this opens a gap at the $\Gamma$ point but leaves the Dirac points intact. For this case, we also did a tight-binding fit and computed a Hartree-Fock phase diagram similar to those in Refs.~[\onlinecite{Ruegg11_2,Ruegg:prb12,Yang:prb11a}] to determine how the gap opening from strain influences the tendency towards realizing interaction-generated topological phases. We find the gap at the $\Gamma$ point suppresses the generation of topological phases at weak coupling (small Hubbard $U$) but leaves the topological phase predictions over the parameter regime relevant to (LaNiO$_3$)$_2$/(LaAlO$_3$)$_N$ (where the Dirac points are relevant) essentially unchanged. The main effect of the externally imposed strain along [001] is to produce an orbital polarization which then favors an antiferromagnetic spin order. This effectively expands the region of the phase diagram for antiferromagnetism. 

Second, within the LSDA+$U$, we identify a breathing distortion of the oxygen cages surrounding the Ni ions. This distortion breaks the inversion symmetry which renders the two Ni-sites on the buckled honeycomb lattice inequivalent. On the basis of an effective tight-binding model we demonstrate that such a distortion competes with a potential topological phase in the fully spin polarized system.

Our paper is organized as follows.  In Sec.~\ref{sec:DFT} we describe our density functional theory calculations for the lattice distorted (LaNiO$_3$)$_2$/(LaAlO$_3$)$_N$ system.  In Sec.~\ref{sec:TB} we present the tight-binding fit to the DFT results, and in Sec.~\ref{sec:HF} we give the Hartee-Fock calculations for the distorted system.  We discuss promising experimental approaches for observing topological phases in oxide heterostructures in Sec.~\ref{sec:exp} and finally conclude in Sec.~\ref{sec:conclusions}.

\section{DFT Results}
\label{sec:DFT}
\subsection{Details of the calculation}
We have studied the electronic structure of the
(LaNiO$_3$)$_2$/(LaAlO$_3$)$_{10}$ supercell (see Fig. 1) using density
functional theory\cite{Hohenberg:pr64,Kohn:pr65} (DFT)  within the local density approximation (LDA)\cite{Kohn:pr65} and the generelized gradient approximation (GGA) with the Perdew-Becke-Erzenhof parametrization, as implemented in the Vienna ab
initio simulation package (VASP).\cite{Kresse:prb96} We used the projector
augmented wave pseudopotentials for all our calculations.\cite{Blochl:prb94}  A
plane-wave cutoff energy of 600 eV and a $6\times 6 \times6$ $k$-point
grid was chosen for integrating over the Brillouin zone. The energies are converged to within 10$^{-6}$ eV/cell and all forces to within 0.004 eV/\AA. The pseudocubic in-plane
lattice constant for the unstrained supercell was chosen as 3.79 \AA\,  which
corresponds to the experimental pseudocubic lattice constant
of bulk LaAlO$_3$. We performed full atomic relaxation and optimized over the out-of-plane lattice constant of the supercell. To treat correlation effects within DFT, we also performed LSDA+$U$ calculations for the fully relaxed system within a simplified rotationally invariant scheme.\cite{Dudarev:1998} We used an effective local interaction parameter $U_{\rm eff}=5.74 \mbox{ eV}$.\cite{Gou:2011,Ruegg:prb12} 

Since the lowest energy configuration of the lattice is one with relatively little change from the ideal configuration (especially as measured by changes in the bandstructure), we also considered a symmetry breaking strain along [001] which does have a significant influence on the band structure.  The LDA band gap for
LaAlO$_3$ in our calculation is $E_g \approx 3.8\mbox{ eV}$ (experimental value $E_g = 5.6 \mbox{eV}$ \cite{Seo:prb11}).
As expected, this wide-band gap leads to a strong confinement
of the electronic degrees of freedom in the LaNiO$_3$ bilayer.\cite{Ruegg:prb12} Physically, then, the role of the LaAlO$_3$ capping layers is simply to provide a ``vacuum" for the LaNiO$_3$ bilayer.  Without the capping layers, the bilayer would not be stable and would be difficult to manipulate experimentally.

Although we are studying a 3D system formed by a supercell with a large period along the [111] direction, the strong quantum confinement of the conduction electrons to the LaNiO$_3$ bilayer implies a negligible $k_z$ dependence of the electronic structure. The system is therefore quasi-2D and we present band structures in the hexagonal 2D Brillouin zone for $k_z=0$.

\subsection{Fully Relaxed System: LDA/GGA}
In order to study lattice relaxation effects, particularly the effect of the the oxygen octahedral tilts within the superlattice,\cite{May:prb10} we compute the electronic structure of a fully relaxed system of the (LaNiO$_3$)$_2$/(LaAlO$_3$)$_{10}$ superlattice. Both LaAlO$_3$ and LaNiO$_3$ have a rhombohedral perovskite structure in the ground state with a tilted AlO$_6$ and NiO$_6$ octahedral network in which alternating octahedral cages undergo rotations about the [111] axis with the same angle but opposite sign, see Fig.~\ref{fig1b} (in the Glazer notation, such a distortion is denoted by $a^-a^-a^-$). At low temperature, the rhombohedral cell angle $\alpha$ [Fig.~\ref{fig1b}] in LaAlO$_3$ and LaNiO$_3$ is found to be 60.1$^\circ$ and 60.8$^\circ$, respectively.\cite{Hayward:prb05,Garcia:prb92} Both these values are very close to cubic symmetry (which would correspond to $\alpha=60^\circ$) and in our calculations we neglected the small deviations from this ideal value. 

The rotations of the octahedral oxygen cages are among the most important lattice distortion effects in perovskites.\cite{Rondinelli:am12} In both LaAlO$_3$ and LaNiO$_3$, the tilts are described by a rotation angle $\phi$ around the cubic [111] direction. Because the direction of rotation alternates between neighboring octahedra [see Fig.~\ref{fig1b}] the idealized cubic unit cell is doubled in bulk. However, the size of the unit cell for the studied [111] superlattice is not affected by the tilts. Experimentally, the tilt angle $\phi$ is found to be 5.74$^\circ$ in LaAlO$_3$\cite{Hayward:prb05} while in LaNiO$_3$ it is found to be 9.2$^\circ$.\cite{Garcia:prb92} In our calculations, we start with a (LaNiO$_3$)$_2$/(LaAlO$_3$)$_{10}$ superlattice with initial octahedral tilt angles $\phi=\pm 6^\circ$ uniformly through out the structure (with opposite signs on neighboring octahedra). We then let the atoms relax to their minimal energy position and optimize over the out-of plane lattice constant of the supercell while fixing the in-plane lattice constant at LaAlO$_3$ (3.79 \AA). The optimized out-of-plane lattice constant was found to be 25.70~\AA\ within LDA and 26.40~\AA\ within GGA. For comparison, the value of an ideal cubic system with $a_0=3.79$ \AA\ would be $4\sqrt{3}a_0=26.26$\ \AA. 
In addition, we have also verified that an initial breathing distortion, where the volume of neighboring oxygen cages alternates, relaxes back to a non-breathing structure.

\begin{figure}[htb]
\includegraphics[width=0.8\linewidth]{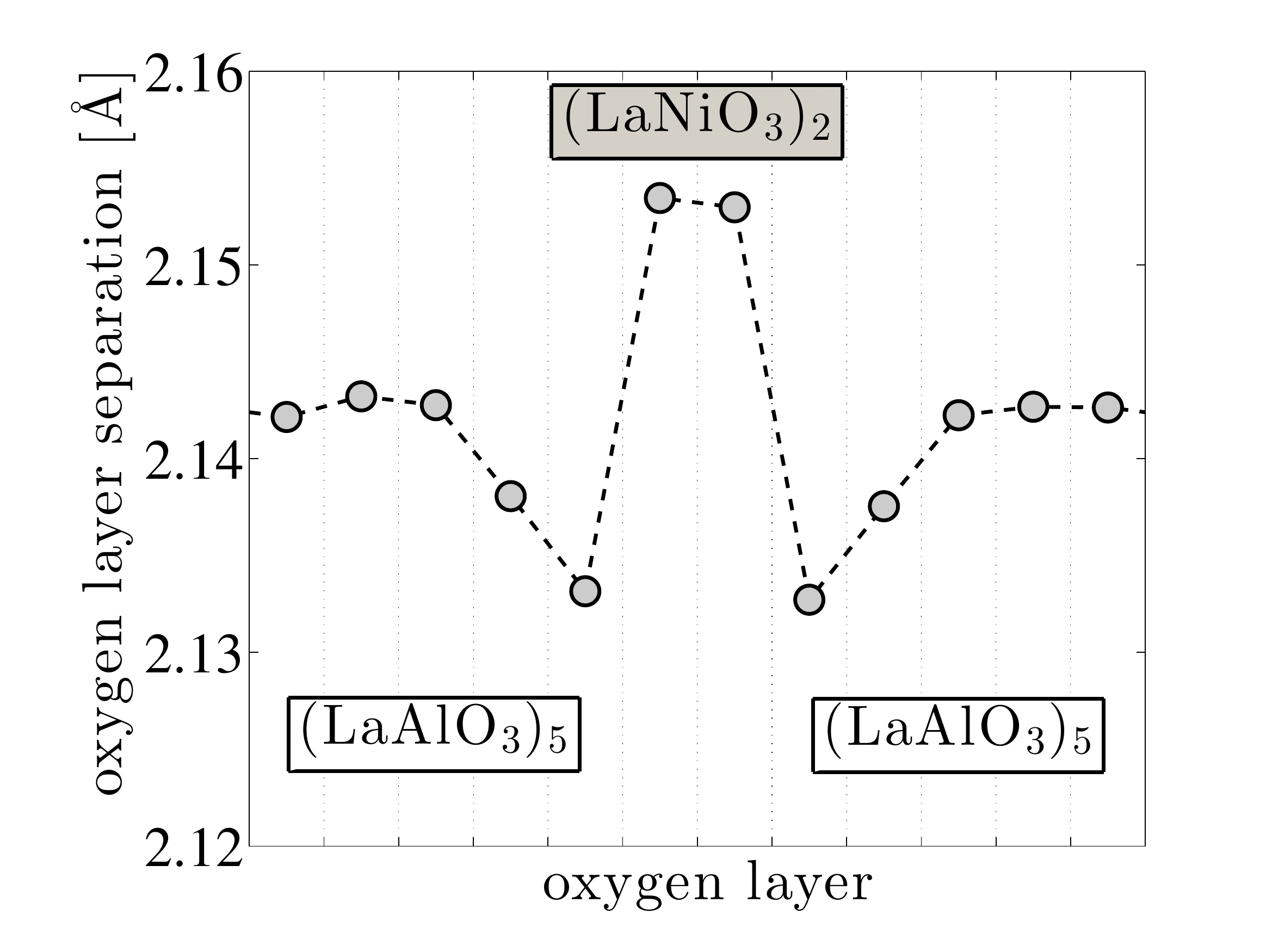}
\caption{(color online) Separation between neighboring oxygen layers in the fully relaxed (LaNiO$_3$)$_2$/(LaAlO$_3$)$_{10}$ superlattice. The results are obtained within the LDA approximation to DFT.}
\label{fig:c_distance}
\end{figure}

The LDA out-of plane relaxation is illustrated in Fig.~\ref{fig:c_distance} where the separation between neighboring oxygen layers is shown. There is a compression near the interface between LaAlO$_3$ and LaNiO$_3$, and an expansion within the LaNiO$_3$ bilayer. This is in accordance with the fact that the pseudocubic lattice constant of LaNiO$_3$ is slightly larger (3.85 \AA) than that of LaAlO$_3$. 

\begin{figure}[htb]
\includegraphics[width=0.8\linewidth]{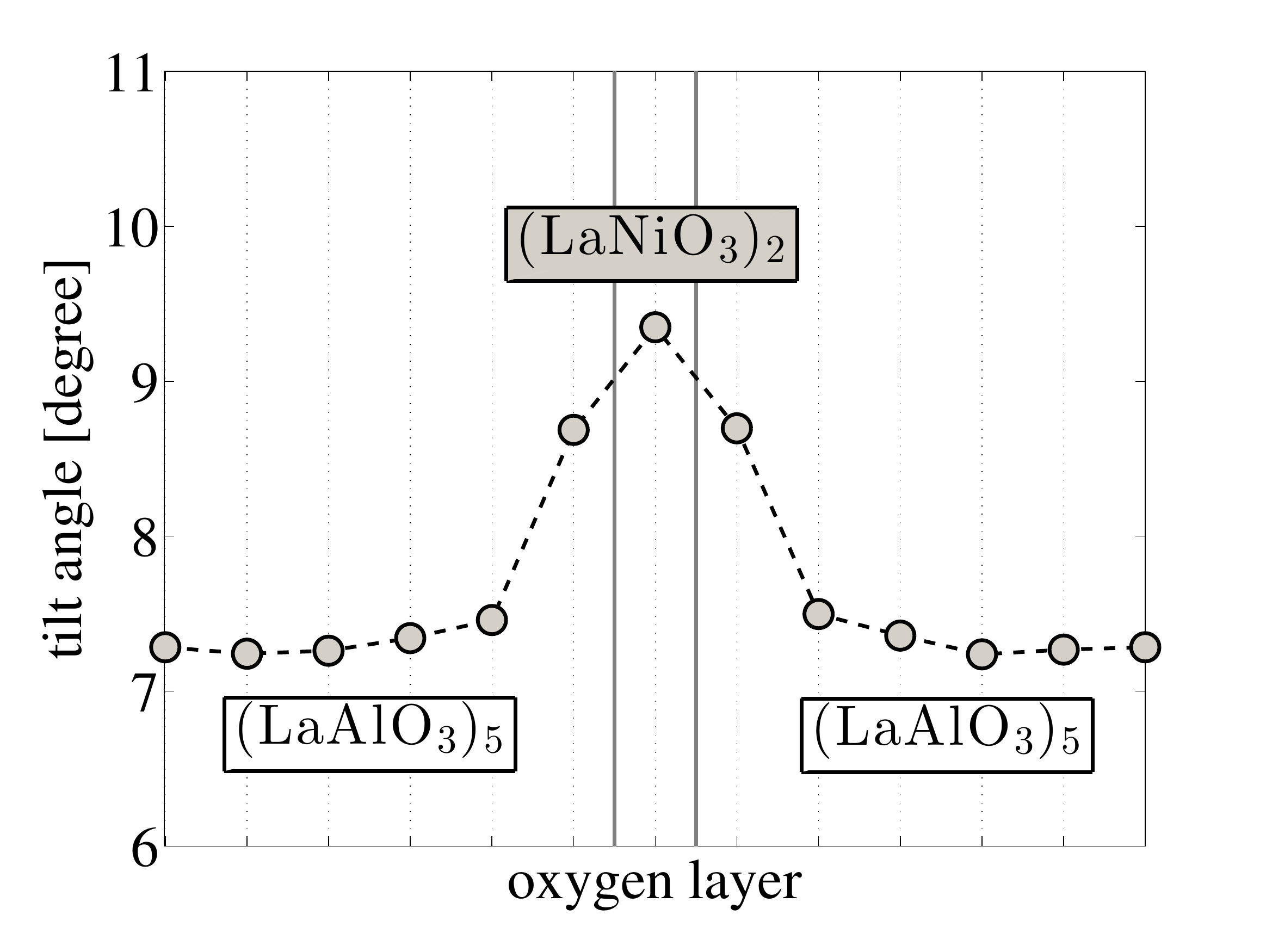}
\caption{Layer resolved octahedral rotation angles for the (LaNiO$_3$)$_2$/(LaAlO$_3$)$_{10}$ supercell.  The results are obtained within the LDA approximation to DFT.  As Fig.~\ref{fig:relaxed_BS} shows, these rotations do not lift the quadratic band touching at the $\Gamma$ point or the Dirac points at K, K' in the Brillouin zone.}
\label{fig:rotations}
\end{figure}

The layer-resolved angle of the rotations of the octahedral oxygen cages as obtained within LDA  is shown in Fig.~\ref{fig:rotations}. It monotonically interpolates between $\phi\approx 7.2^\circ$ in LaAlO$_3$ and $\phi\approx 9.3^\circ$ in LaNiO$_3$ for the LDA calculation. While the computed value for LaNiO$_3$ is very close to the experimental bulk value ($9.2^\circ$), the rotation angle in LaAlO$_3$ is slightly higher than both the experimental ($5.7^\circ$) and the LDA value obtained in Ref.~\onlinecite{Seo:prb11} ($6.1^\circ$). However, Ref.~\onlinecite{Seo:prb11} uses an LDA optimized smaller lattice constant (3.74~\AA) to better match the experimental value. Because we are using the actual experimental lattice constant (3.79~\AA), the tilt angles are slightly overestimated by our LDA calculations.

\begin{figure}[htb]
\includegraphics[width=1\linewidth]{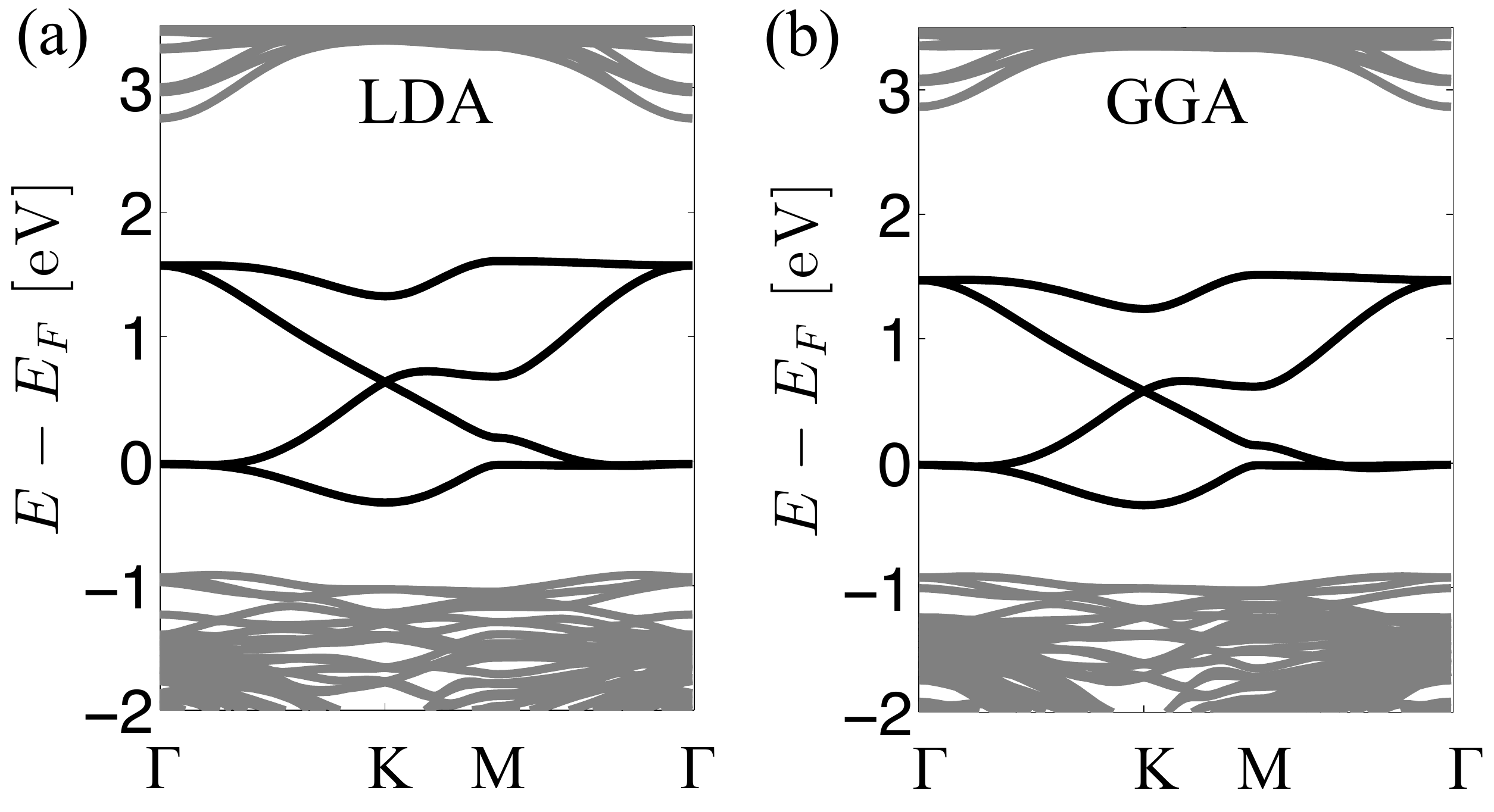}
\caption{DFT band structure of the fully relaxed (LaNiO$_3$)$_2$/(LaAlO$_3$)$_{10}$ system within (a) LDA and (b) GGA. Crucially, the quadratic band touching at the $\Gamma$ point as well as the linear crossings at K and K$'$ are preserved.
}
\label{fig:relaxed_BS}
\end{figure}

Our results for the fully relaxed band structure in the quasi 2D Brillouin zone are shown in Fig.~\ref{fig:relaxed_BS}. Overall, we find little variation between LDA and GGA and the band structure is close to the band structure of the unrelaxed system discussed previously.\cite{Ruegg:prb12} Moreover, both the quadratic band touching at the $\Gamma$ point as well as the linear band crossing at the K and K$'$ points in the unrelaxed system are preserved in the fully relaxed system. As a result, earlier predictions for topological phases based on an interaction-induced gap opening at the $\Gamma$ point (or at the Dirac points in a fully spin-polarized system) are essentially unaffected by the lattice relaxation.\cite{Ruegg11_2,Ruegg:prb12,Yang:prb11a} However, fully relaxing the structure results in two quantitative changes as compared to the ideal structure.\cite{Ruegg:prb12} First, there is an overall reduction of the $e_g$ band width. Second, the band gap of the LaAlO$_3$ increases by about 0.8~eV. We found that both changes are predominantly because of the rotations of the oxygen octahedra.

The robustness of the touching points are understood from symmetry considerations. Both the ideal and the relaxed structure have the trigonal point group symmetry $D_{3d}$ with a $C_3$ axis along the [111] direction and three $C_2'$ axes in the plane perpendicular to [111]. The $C_2'$ axes lie in the oxygen layer which is sandwiched between the two Ni layers. $D_{3d}$ permits two 2D irreducible representations and hence the two-fold degeneracies at $\Gamma$ are protected by the trigonal symmetry. Furthermore, under inversion symmetry, $k_x\rightarrow -k_x$, $k_y\rightarrow -k_y$ and $k_z\rightarrow -k_z$ so that the dispersion for $k_z=0$ is inversion symmetric in the 2D Brillouin zone, too. Hence, the quadratic band touching point is stable: it does not split into four Dirac points under symmetry preserving perturbations. The symmetry group for the K and K' points is $C_{3v}$ which also permits a two-dimensional irreducible representation. Thus, the Dirac points at half filling are protected by the trigonal symmetry as well.
Note the difference to Bernal stacked bilayer graphene which also has the $D_{3d}$ point group symmetry.\cite{McCann:2013} There, as opposed to the LaNiO$_3$ bilayer, the quadratic band touching point occurs at the K and K' points whose little group is $C_{3v}$ and allows for the splitting into four Dirac points without breaking the symmetry.\cite{McCann:2013}

\subsection{Fully Relaxed System: LSDA+$U$}
\label{sec:LSDAU}
We now briefly discuss correlation effects within the LSAD+$U$ scheme. For bulk nickelates, it is known that LDA/GGA+$U$ wrongly predicts a ferromagnetic ground state not seen in experiments. For example, LaNiO$_3$ is a paramagnetic metal but a ferromagnetic ground state is found for a typical value of $U\sim 6$~eV.\cite{Gou:2011} A similar problem occurs for insulating nickelates such as LuNiO$_3$ which are antiferromagnetic in the low temperature regime but, again, DFT+$U$ (and even DFT+DMFT\cite{Park:2012}) predicts a ferromagnetic ground state. Although the magnetism in our [111] sandwich structure can differ from bulk, in view of these problems known for the bulk systems, the DFT+$U$ predictions for the magnetism should be taken with caution.

Performing the LSDA+$U$ calculations for the fully relaxed [111] sandwich structure, we find a fully polarized ferromagnetic ground state, similar to the ideal system discussed previousely.\cite{Ruegg:prb12} What is interesting about the scenario of a fully polarized ferromagnet is that the Fermi energy is placed right at the Dirac points of the majority band. This opens the possibility for unusual interaction-driven phases, as discuss previously.\cite{Yang:prb11a,Ruegg:prb12} 

\begin{figure}
\includegraphics[width=0.8\linewidth]{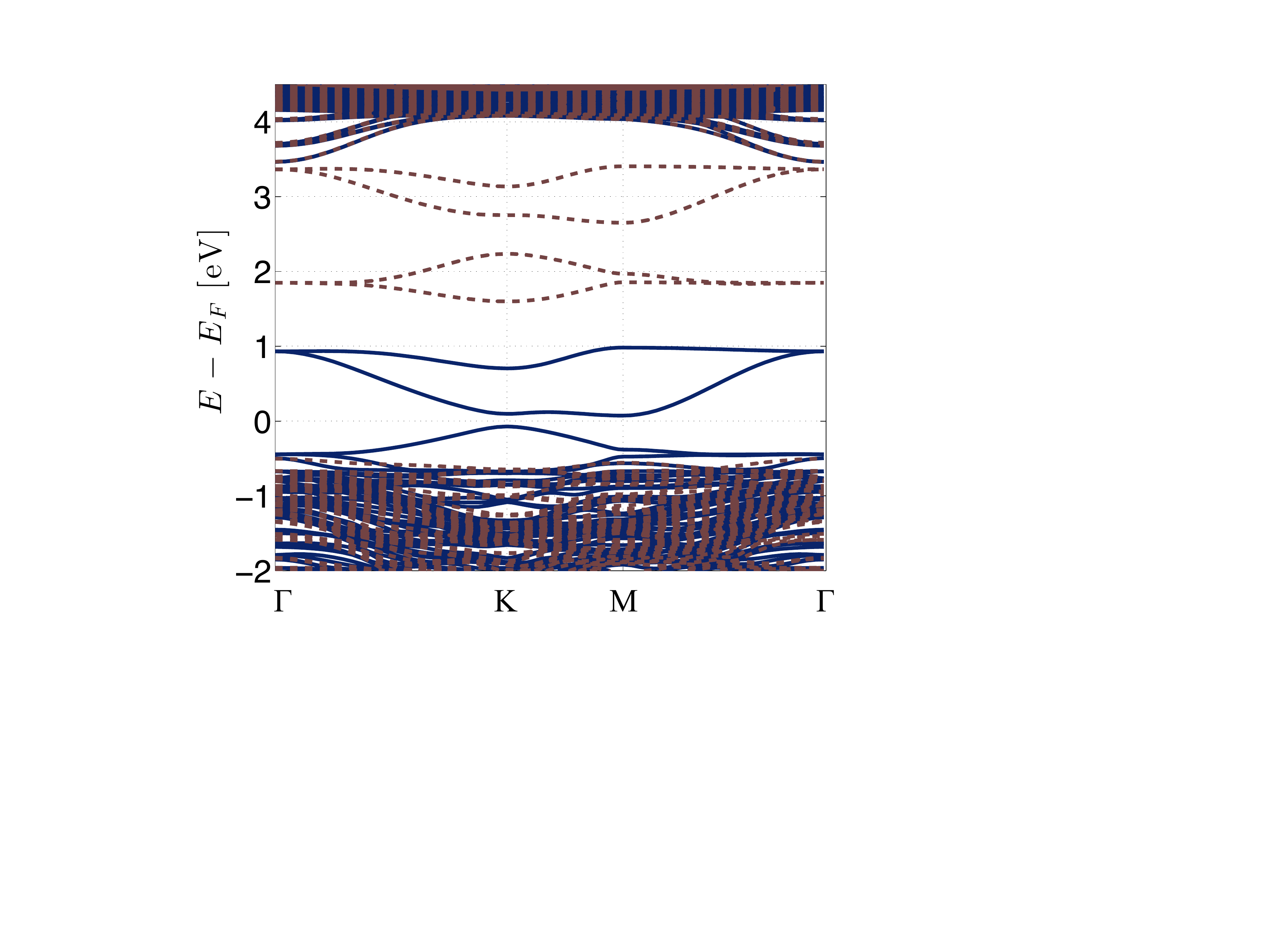}
\caption{The band structure of the fully relaxed (LaNiO$_3$)$_2$/(LaAlO$_3$)$_{10}$ system within the LSDA+$U$ which predicts a fully polarized ferromagnetic ground state. A gap opens at the Dirac points as a consequence of a breathing distortion which reduces the volume of the oxygen cage on one sublattice while it increases it on the other sublattice. Solid lines are the majority, dashed lines the minority bands.}
\label{fig:LSDAU}
\end{figure}

As opposed to the LDA, we find that the LSDA+$U$ sustains a breathing distortion where the volumes of the oxygen cages in the first Ni layer are reduced as compared to the volumes of the octahedra in the second Ni layer. This distortion breaks the inversion symmetry which opens a gap at the Dirac points. The resulting band structure is shown in Fig.~\ref{fig:LSDAU}. The two Ni atoms now have different magnetic moments amounting to 0.96$\mu_B$ and $1.24\mu_B$, respectively ($\mu_B$ is the Bohr magneton), with the larger moment being surrounded by the larger oxygen cage. Interestingly, the difference in the total number of $d$ electrons between the two Ni sites is very small and amounts to roughly 0.01. We also note that the orbital polarization is vanishingly small. 
The possibility that a breathing distortion is stabilized in insulating rare-earth nickelates has been considered previously.\cite{Mizokawa:2000,Gou:2011,Park:2012} It is also a subject of great interest in thin films, where strain and quantum confinement may favor such a distortion for LaNiO$_3$ (which is metallic and undistorted in bulk).\cite{Chakhalian:prl11,Freeland:2011} In Sec.~\ref{sec:BO} we argue that such a breathing distortion can be modeled as a bond-order wave in an effective model which only includes the $e_g$ orbitals. This provides an alternative perspective to the commonly used one which treats the oxygens explicitly.\cite{Mizokawa:2000,Gou:2011,Park:2012} In addition, on the basis of this effective model, we also argue that such a breathing distortion would in general compete with a topological phase predicted to occur in the ferromagnetic phase for certain parameters within Hartree-Fock.

On the other hand, to the best of our knowledge, there are presently no experimental indications for inequivalent Ni sites\cite{Middey:apl12} and the experimental relevance of the theoretical observation of a breathing distortion is currently unclear. In view of this uncertainty, we postpone a detailed study of the interplay between structural distortion and correlation effects. In Sec.~\ref{sec:HF}, instead, we again discuss the possible role of interactions using the Hartree-Fock approximation for an effective multi-orbital model. This approach allows us to map out a larger parameter regime (which also includes external strain) and to access a larger range of possible phases.

\subsection{Symmetry-breaking Strain along [001]: LDA}

The main conclusion of our DFT calculations for the fully relaxed system is that the crucial band features favoring interaction-driven topological phases at the Hartree-Fock level remain intact. In this section, we invert the question - What type of lattice distortion {\em can} open a gap in the band structure at $\Gamma$ (and the Dirac points in the fully polarized system) and thereby compete with possible topological phases?

\begin{figure}[htb]
\includegraphics[width=1\linewidth]{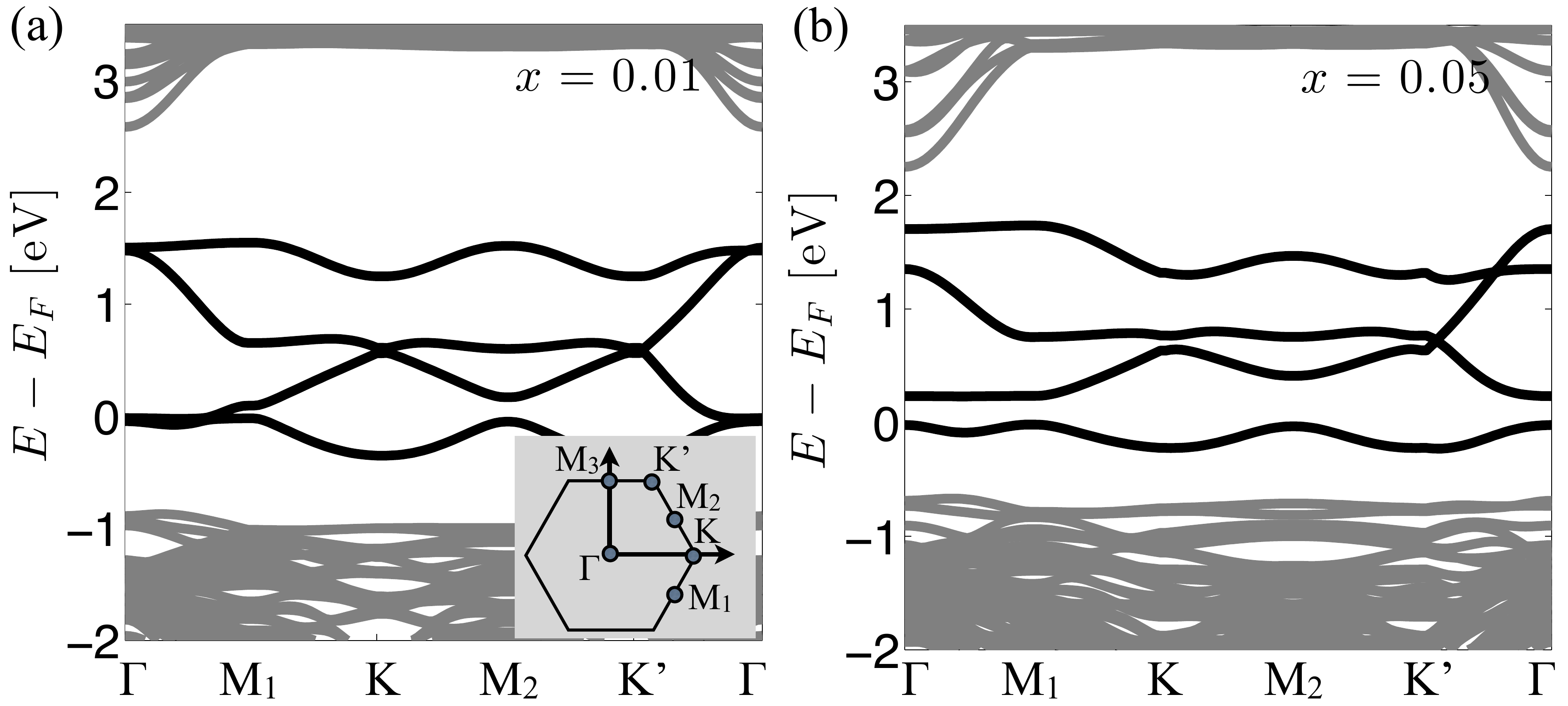}
\caption{The band structure of the [001] strained (see text) for (a) $x=0.01$ and (b) $x=0.05$. Note that a strain of $x=0.05$ opens a sizable bandgap at the $\Gamma$ point but a Dirac point remains close to K$'$. The inset shows the special points in the Brillouin zone.}
\label{fig:strain_BS}
\end{figure}

We find that we are able to open a gap at the $\Gamma$ point with a strain applied along the [001] direction (though the Dirac point remains for the studied range of strain), which breaks the rotational symmetry about the [111] direction. Specifically, we {\em impose} the following lattice strain on the system and then compute the resulting band structure:
\begin{eqnarray}
{\bf a_1}&=&a_0(1-\mu x)\bf i,\nonumber\\
{\bf a_2}&=&a_0(1-\mu x)\bf j, \nonumber \\
{\bf a_3}&=&a_0(1+x)\bf k,
\label{eq:strain}
\end{eqnarray}
where ${\bf a_i}$ are the lattice vectors under strain in the three cubic directions, ${\bf i,j,k}$, $a_0=3.79$~\AA\ is the undistorted lattice constant, $x$ is the fraction of the lattice lengthening or contraction in the $\bf k$-direction, and $\mu=0.24$ is the Poisson's ratio for LaAlO$_3$.\cite{Luo:2008} The band structures for $x=0.01$ and $x=0.05$ are shown in Fig.~\ref{fig:strain_BS}. Because strain along the [001] direction breaks the trigonal symmetry, we show both K and K$'$ points as well as two different M points. In terms of the reciprocal lattice vectors ${\bs G}_1$ and ${\bs G}_2$, these points are given by M$_1={\bs G}_1/2$, K$=(2{\bs G}_1+{\bs G}_2)/3$, M$_2=({\bs G}_1+{\bs G}_2)/2$ and K$'=({\bs G}_1+2{\bs G}_2)/3$.

\begin{figure}[htb]
\includegraphics[width=0.49\linewidth]{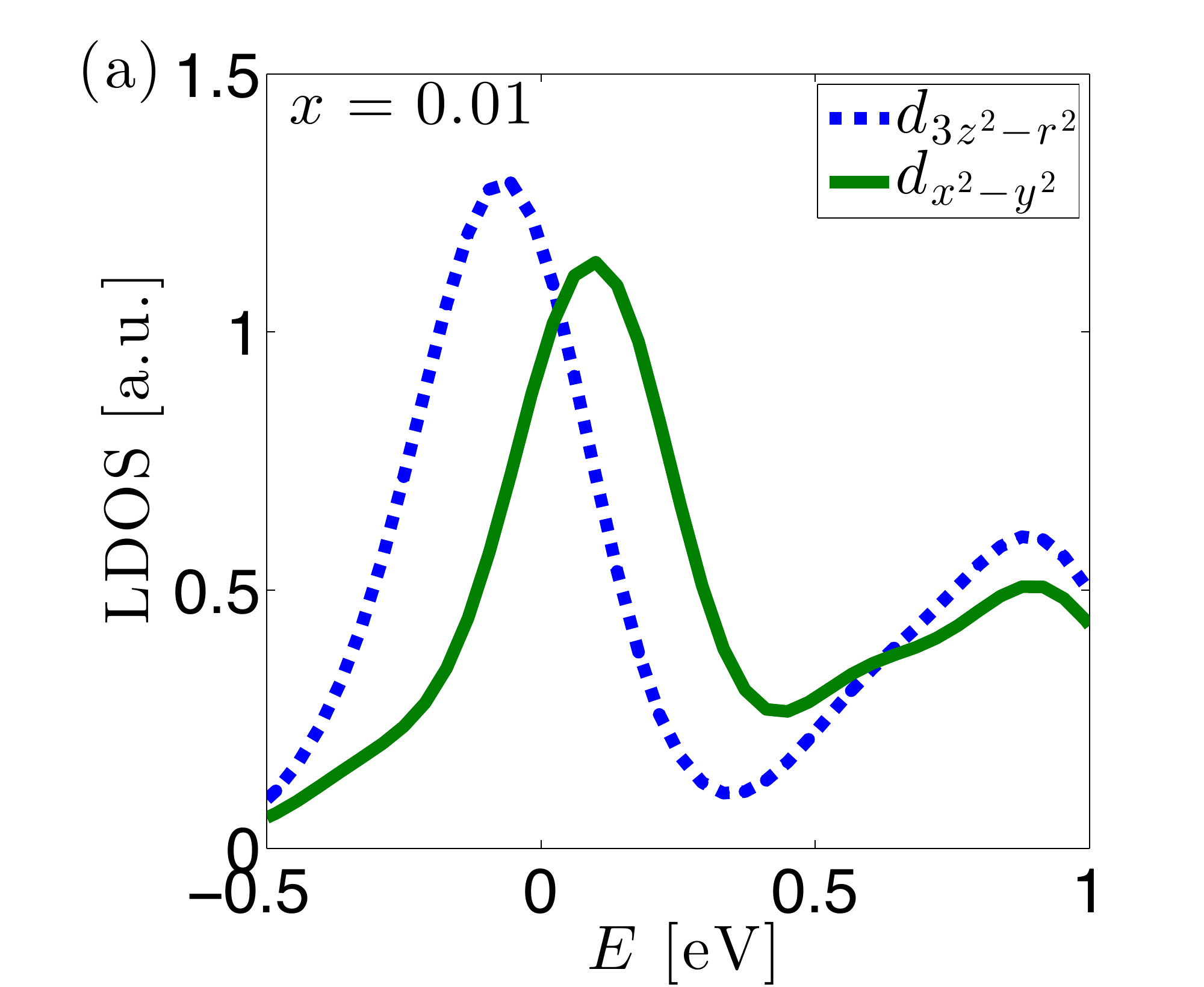}
\includegraphics[width=0.49\linewidth]{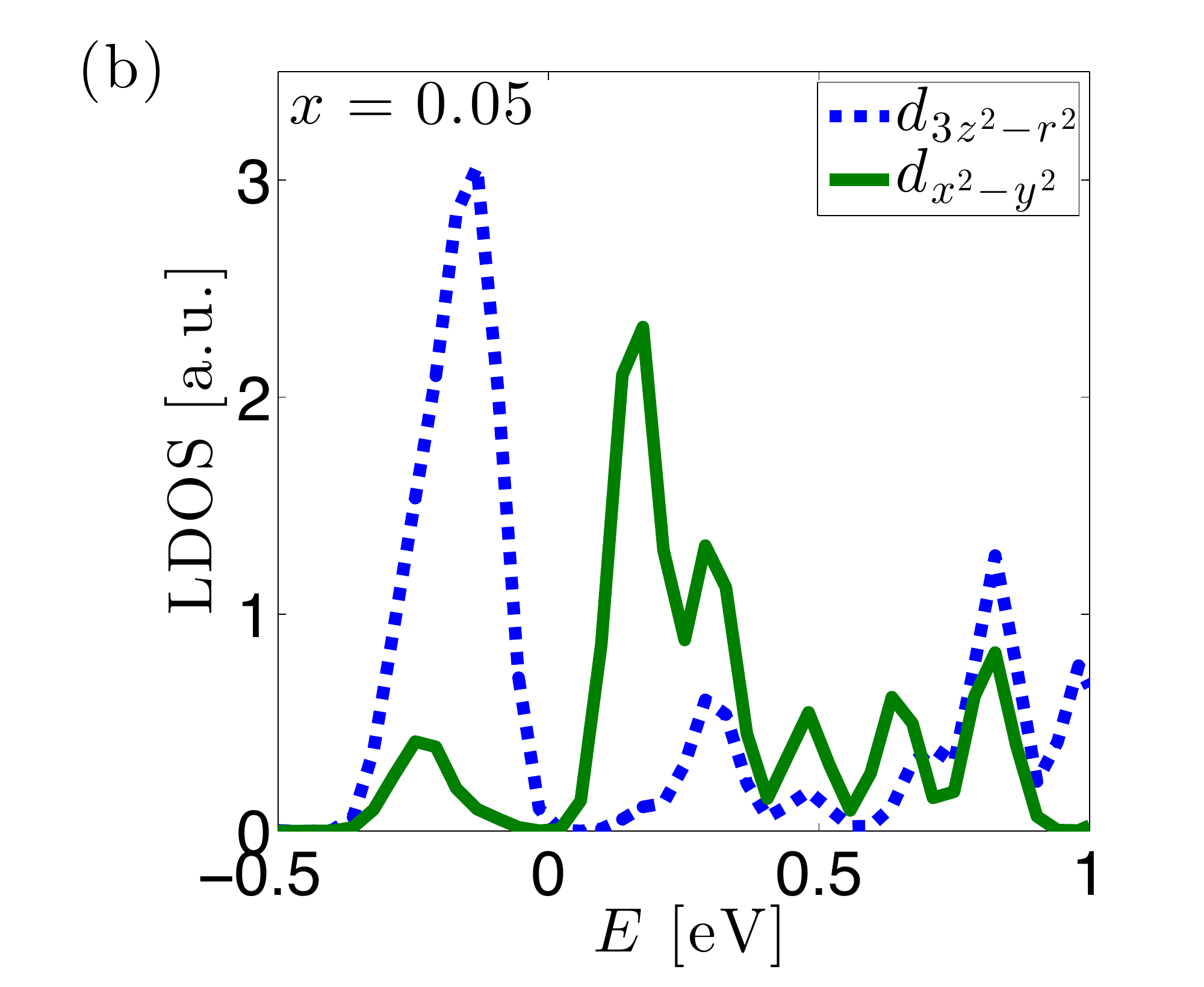}
\caption{(color online) The orbital-resolved local density of states (LDOS) for a [001] strained (LaNiO$_3$)$_2$/(LaAlO$_3$)$_{10}$ system near the Fermi energy $E=0$ for (a) $x=0.01$ and (b) $x=0.05$. Only the dominant contribution from the $e_g$ orbitals is shown. We find no symmetry breaking for the two Ni sites in terms of the LDOS, which are numerically identical. The strain induces an orbital splitting and significant orbital polarization of the $e_g$ orbitals.}
\label{fig:strain_LDOS}
\end{figure}

In Fig.~\ref{fig:strain_LDOS} we plot the orbital-resolved local density of states (LDOS) for a [001] strained LaNiO$_3$)$_2$/(LaAlO$_3$)$_{10}$ system with (a) $x=0.01$ and (b) $x=0.05$. We find no symmetry breaking in the LDOS for the two Ni sites, suggesting they are equivalent sites under the [001] strain. The external strain splits the energy of the two $e_g$ orbitals and induces an orbital polarization. For the strongly strained case with $x=0.05$ a gap in the LDOS at the Fermi level is noticeable, which is consistent with the gap at 1/4 filled for {\em spin unpolarized} $e_g$ bands in Fig.~\ref{fig:strain_BS}. It would be interesting to see if this large orbital polarization persists in a more careful treatment using strongly correlated methods that explicitly include the oxygen orbitals,\cite{Han:prl11} as to the best of our knowledge this issue has not been addressed in the [111] growth direction.

\section{Effective Tight-binding Models}
\label{sec:TB}

For comparison among the unrelaxed, fully relaxed, and [001] strained systems, we perform a tight-binding fit to both the fully relaxed system and the [001] strained system.  We find there is little difference in the fitting parameters between the fully relaxed system and the undistorted system the authors previously studied,\cite{Ruegg:prb12} while there is a significant difference for the strained system in which a gap opens at the $\Gamma$ point. In Sec.~\ref{sec:HF} we will compute a Hartree-Fock phase diagram for the tight-binding model with interactions for the case of strain along [001].

\subsection{Fitting for the Fully Relaxed System}

We begin with the fully relaxed system for which the quadratic band touching at the $\Gamma$ point is preserved.  Following Ref.~[\onlinecite{Ruegg:prb12}], we consider a tight-binding model based only on the nickel $e_g$ orbitals that includes nearest-neighbor hopping via the oxygen $p$-orbitals and also second-neighbor hopping via the oxygen $p$-orbitals.  We find a better fit can be obtained by including the small differences in the hopping to ``outer" versus ``inner" oxygen atoms.\cite{Ruegg:prb12}  Assuming trigonal symmetry is preserved (a result consistent with our fully relaxed DFT results), we take the nearest-neighbor Slater-Koster parameters for hopping along the $z$-direction to be described by the matrix
\begin{equation}
\hat{t}_z=-\begin{pmatrix}
t&0\\
0&t_{\delta}
\end{pmatrix}
\label{eq:tz}
\end{equation}
in the basis $(d_{z^2},d_{x^2-y^2})$.  Here $t$ includes predominantly the hopping via the intermediate oxygen while $t_{\delta}$ arises from the direct overlap and is small. We set $t_{\delta}=0$ in the following. Assuming that the nearest-neighbor hopping in the $x$ and $y$ directions are equivalent to the hopping along the $z$ direction, we obtain the corresponding matrices by a rotation of the $e_g$-orbitals around [111] by $\pm 2\pi/3$. The matrix for the rotation by $2\pi/3$ is
\begin{equation}
\hat{R}=\begin{pmatrix}
-1/2&\sqrt{3}/2\\
-\sqrt{3}/2&-1/2
\end{pmatrix}.
\label{eq:R}
\end{equation}
As a result, we find
\begin{equation}
\hat{t}_x=\hat{R}^T\hat{t}_z\hat{R}, \quad \hat{t}_y=\hat{R}^T\hat{t}_x\hat{R}.
\end{equation}

The Slater-Koster parameters for second-neighbor hopping via two intermediate oxygen atoms define the matrix
\begin{equation}
\hat{t}_{xy}=-\begin{pmatrix}
t'/2&\sqrt{3}\Delta/2\\
-\sqrt{3}\Delta/2&-3t'/2
\end{pmatrix}.
\label{eq:txy}
\end{equation}
The parameters take into account the lowest-order processes for second-neighbor hopping. The off-diagonal entries proportional to $\Delta$ are allowed in the bilayer system discussed here (as opposed to a perfect cubic system) because the two possible paths connecting second-neighbor transition-metal ions are not equivalent: they either involve ``inner" or ``outer" oxygens.\cite{Ruegg:prb12} Note that $\hat t_{xy}$ is not symmetric if $\Delta\neq 0$ which means that there is an associated direction for the hopping. We use the convention that $\hat t_{xy}$ denotes the hopping of an electron along a second neighbor bond which is reached by first following the $y$-axis and then the $x$-axis of the cube.
By rotating the orbitals, we also obtain the second-neighbor hopping along the other directions:
\begin{equation}
\hat{t}_{yz}=\hat{R}^T\hat{t}_{xy}\hat{R},\quad \hat{t}_{zx}=\hat{R}^T\hat{t}_{yz}\hat{R}.
\end{equation}
Including the above introduced hopping matrices, the generalized tight-binding model now takes the form
\begin{eqnarray}
H_0&=&\sum_{{\bs r}\in A}\sum_{s}\sum_{u=xyz}\left(\vec{d}^{\dag}_{s,{\bs r}}\hat{t}_u\vec{d}_{s,{\bs r}+{\bs e}_u}+{\rm h.c.}\right)\nonumber\\
&+&\sum_{{\bs r}\in A}\sum_{s}\sum_{u=xyz}\left(\vec{d}^{\dag}_{s,{\bs r}}\hat{t}_{u,u+1}\vec{d}_{s,{\bs r}+{\bs e}_u-{\bs e}_{u+1}}+{\rm h.c.}\right)\label{eq:H0}\\
&+&\sum_{{\bs r}\in B}\sum_{s}\sum_{u=xyz}\left(\vec{d}^{\dag}_{s,{\bs r}}\hat{t}_{u,u+1}\vec{d}_{s,{\bs r}-{\bs e}_u+{\bs e}_{u+1}}+{\rm h.c.}\right).\nonumber
\label{eq:H0}
\end{eqnarray}
Here, $\vec{d}_{s}=(d_{z^2,s},d_{x^2-y^2,s})^T$ is a vector in orbital space, $s=\uparrow$, $\downarrow$ is the spin and the notation $u+1$ refers to $y$ if $u=x$ with a cyclic extension to the other elements. 

\begin{table}
\begin{ruledtabular}
\begin{tabular}{l | l l l l l}
fit & $t$ [eV]& $t'$ [eV]& $\Delta$ [eV] & $E_F$ [eV]\\
\hline
unrelaxed (LDA)\cite{Ruegg:prb12}& 0.598 & 0.062 & -0.023 &  -0.693\\
fully relaxed (LDA)& 0.541 & 0.045 & -0.017 &  -0.641\\
fully relaxed (GGA)& 0.508 & 0.046 & -0.016 &  -0.593
\end{tabular}
\end{ruledtabular}
\caption{Parameters obtained in tight-binding fits to the $e_g$ DFT band structure of the unrelaxed and fully relaxed 12 layer superlattice shown in Fig.~ \ref{fig:relaxed_TB_fit}.}
\label{tab:parameters}
\end{table}

\begin{figure}[htb]
\includegraphics[width=0.8\linewidth]{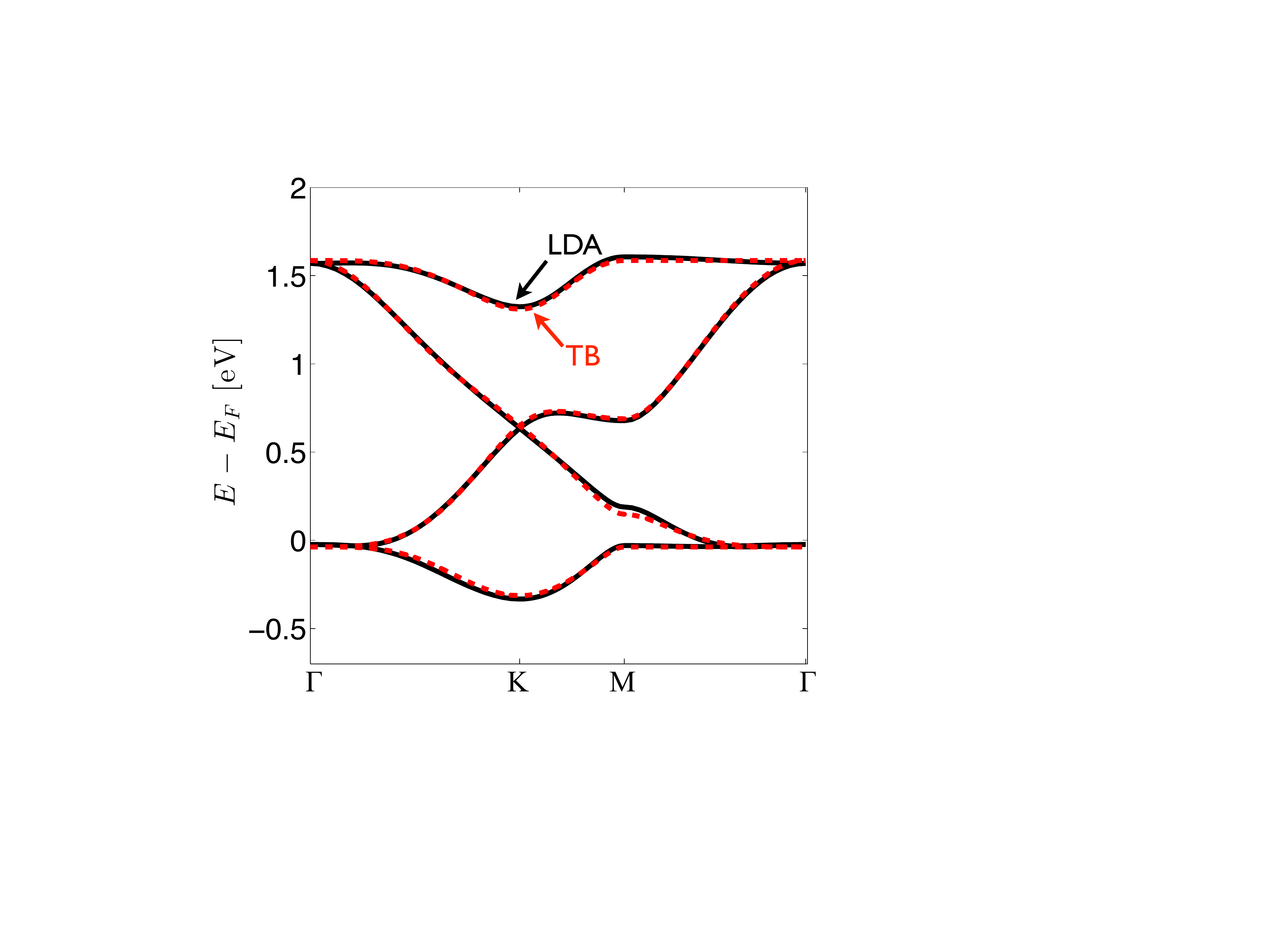}
\caption{(color online) Fully relaxed LDA band structure and tight-binding (TB) fit. A comparison of the tight-binding parameters with those for the unrelaxed system is given in Table~\ref{tab:parameters}.}
\label{fig:relaxed_TB_fit}
\end{figure}

Using the tight-binding model $H_0$ with parameters $t$, $t'$, and $\Delta$ (with $t_\delta=0$), we fitted both the LDA and GGA band structures of the fully relaxed system near the Fermi level. The fitting parameters are listed in Tab.~\ref{tab:parameters} and Fig.~\ref{fig:relaxed_TB_fit} shows the LDA together with the tight-binding band structure for the best fit. As mentioned in the previous section, the relaxation of the lattice including the oxygen tilts does not affect the band structure in a qualitative way and the model $H_0$ captures well the DFT results. Compared to the unrelaxed case there is an overall reduction of the kinetic energy scale by about 10-15\% (see Tab.~\ref{tab:parameters}) which makes the system more susceptible to interaction effects. The phase diagrams in Fig.~2 of Ref.~[\onlinecite{Ruegg11_2}] and in Fig.~7 of Ref.~[\onlinecite{Ruegg:prb12}] will thus have a numerically small shift in the boundaries between different phases for the fully relaxed system. We have explicitly verified this.

\subsection{Fitting for the System with Strain along [001]} 
We now turn to the system with an external strain imposed along the [001] direction. On the level of the tight-binding model, such a distortion introduces several symmetry breaking perturbations in the Hamiltonian.\cite{Nanda:2010,Baena:2011} The most important one is an orbital dependent local energy splitting
\begin{equation}
H_z=\alpha_z\sum_{\bs r}\left(n_{{\bs r},x^2-y^2}-n_{{\bs r},z^2}\right).
\label{eq:Hz}
\end{equation}
For $x>0$ in Eq.~\eqref{eq:strain}, $\alpha_z>0$ and the local energy of the $d_{z^2}$-orbital is lowered as compared to the $d_{x^2-y^2}$-orbital, see Fig.~\ref{fig:strain_LDOS}. Physically, the uniaxial strain elongates the oxygen octahedra along the [001] direction, which results in an orbital field. Besides the orbital field given by $H_z$, strain along [001] also modifies the overlapping matrices thereby inducing an anisotropy in the hopping amplitudes. Hence, in the externally strained case, the form of $H_0$ in Eq.~\eqref{eq:H0} is altered and we introduce the parameter $\eta$ which rescales the nearest-neighbor hopping along the $z$-direction
\begin{equation}
\tilde{t}_z=\eta^{-1} \hat{t}_z,\quad \tilde{t}_x=\hat{t}_x,\quad \tilde{t}_y=\hat{t}_y.
\end{equation}
In contrast to the anisotropy of the nearest-neighbor hopping matrices, we have found that the anisotropy of the second-neighbor hopping matrices is small and does not improve the tight-binding fit in an essential way. In the following, we only keep the hopping anisotropy in the first-neighbor hopping. The anisotropic hopping Hamiltonian is denoted by $\tilde{H}_0$ and the full tight-binding model for the system with strain along [001] is given by
\begin{equation}
H_{\rm strained}=\tilde{H}_0+H_z.
\label{eq:Hstrain}
\end{equation}

\begin{figure}[htb]
\includegraphics[width=1\linewidth]{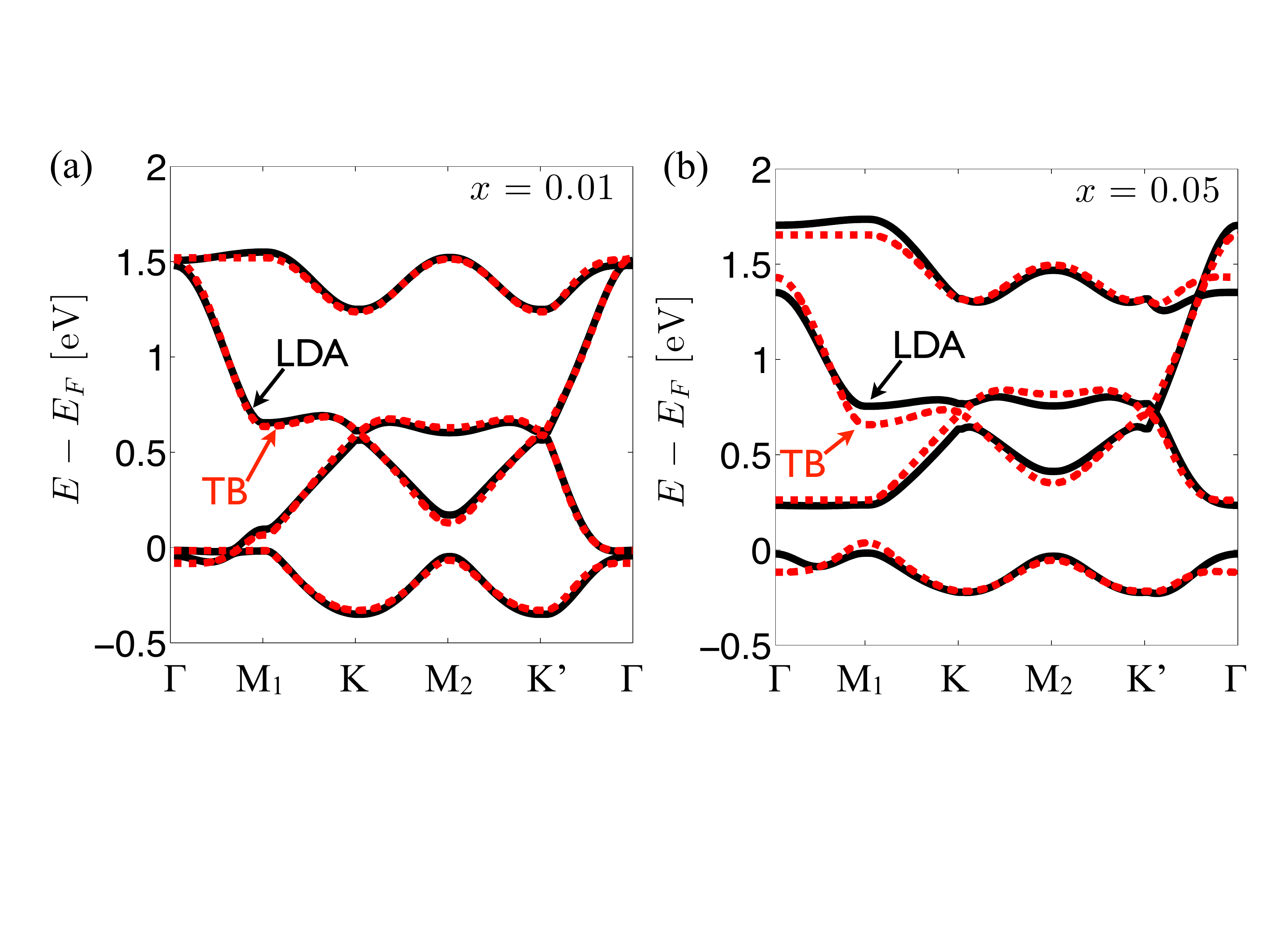}
\caption{(color online) Tight-binding (TB) band structure fits for (a) $x=0.01$ and (b) $x=0.05$ using the model Eq.~\eqref{eq:Hstrain} which includes the 6 parameters $t$, $t'$, $\Delta$, $\alpha_z$, $\eta$ and the Fermi energy $E_F$, see text.}
\label{fig:strained_TB_fit}
\end{figure}

We used the model Eq.~\eqref{eq:Hstrain} to fit the LDA band structure for the case of $x=0.01$ and $x=0.05$ with $x$ given in Eq.~\eqref{eq:strain}. The results are shown in Fig.~\ref{fig:strained_TB_fit}. Overall, the quality of the fit is less good compared to the fully relaxed case, see Fig.~\ref{fig:relaxed_TB_fit}. However, the simple model Eq.~\eqref{eq:Hstrain} correctly captures the overall features of the band structure including the opening of the gap at the $\Gamma$ point for $x=0.05$. The values of the strain induced parameters are $\alpha_z\approx 0.02$ eV and $\eta\approx0.95$ for $x=0.01$ while $\alpha_z\approx0.15$~eV and $\eta=0.86$ for $x=0.05$.

As discussed above, moderate external strain opens a gap at the $\Gamma$ point resulting in an insulating phase for the spin unpolarized system. However, the strain does not open a gap at the Dirac points which are relevant in the fully spin-polarized FM phase. This fact is best understood by linearizing the strained Hamiltonian around K and K' which results in
\begin{equation}
H({\bs k})=v_x(k_x-A_x\tau_z)\sigma_x\tau_z+v_yk_y\sigma_y+\epsilon_0({\bs k})\sigma_0\tau_z
\label{eq:HDirac_strain}
\end{equation}
In the lowest order, the strain-induced perturbations to the ideal Dirac model enter via the following parameters
\begin{eqnarray*}
(v_x,v_y)&=&\left[\frac{3t(1+8t'\alpha_z/t^2)}{4},\frac{3t(1-8t'\alpha_z/t^2)}{4}\right]\\
A_x&=&\frac{2(1-\eta^{-1})}{3t}\\
\epsilon_0({\bs k})&=&-\alpha_zk_x.
\end{eqnarray*}
The external strain along [001] has three effects: (i) it introduces an anisotropy in the Fermi velocity $v_x\neq v_y$, (ii) it shifts the Dirac point along the $k_x$ direction by $A_x$ and (iii) leads to a tilt of the Dirac cones around the $k_y$ axis with opposite tilt angle for K and K' as described by the term $\epsilon_0({\bs k})$. Crucially, however, the external strain does not open a gap at the Dirac points in lowest order. 

\subsection{Breathing distortion as a bond order wave}
\label{sec:BO}
\begin{figure}
\includegraphics[width=0.95\linewidth]{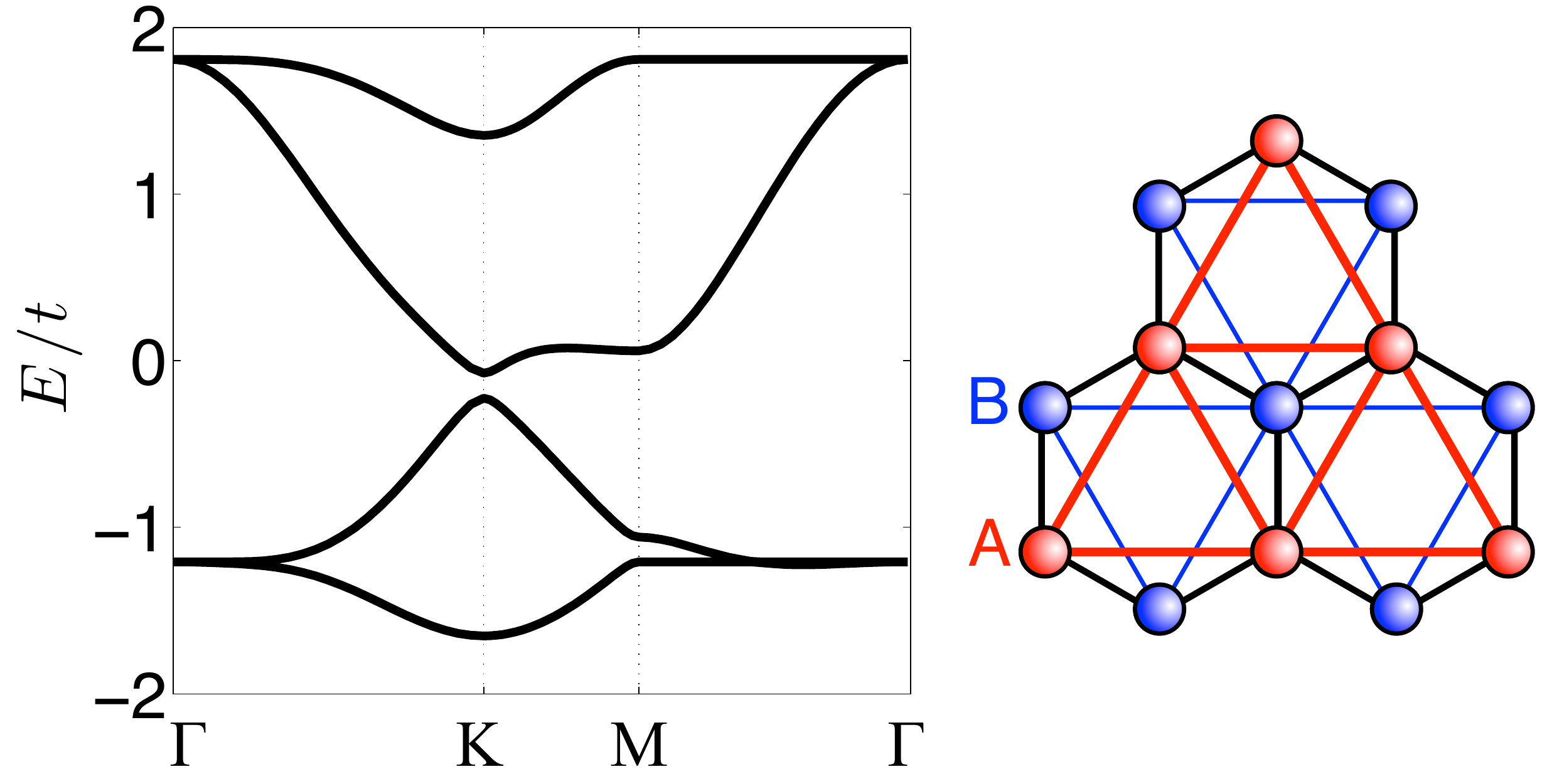}
\caption{Gap opening at the Dirac point due to a bond-order wave in the second-neighbor hopping amplitudes $t'\rightarrow t'\pm\epsilon$. Such a bond-order wave mimics the breathing distortion found in the LSDA+$U$ calculation of Sec.~\ref{sec:LSDAU}. Parameters are $t'=0.1t$ and $\epsilon=0.05t$.}
\label{fig:bond_order}
\end{figure}

Before we proceed with the Hartree-Fock calculations for the [001] strained systems, we would like to provide a perspective of the breathing distortion found in the LSDA+$U$ within the effective tight-binding model Eq.~\eqref{eq:H0}. Because the difference in the charge on the two Ni sites is very small, the term which opens the gap at the Dirac points in the spectrum is not simply a staggered sublattice potential. Instead, an additional perturbation allowed by symmetry can be considered. This is a term which enhances the hopping on sublattice $A$, $t'\rightarrow t'+\epsilon$, while it reduces the hopping on sublattice $B$, $t'\rightarrow t'-\epsilon$, see Fig.~\ref{fig:bond_order}. Such a bond-order wave in the {\it second-neighbor} hopping amplitudes also breaks the inversion symmetry but it leaves the charge on the two Ni sites almost unaffected. The microscopic origin of such a perturbation can be understood from considering the second-neighbor hopping processes via two intermediate oxygens:\cite{Ruegg:prb12} the breathing distortions renders the second-neighbor hopping among two $A$ sites inequivalent from the hopping among two $B$ sites. In Fig.~\ref{fig:bond_order} we show the tight-binding band structure for $t'=0.1t$ and $\epsilon=0.05t$ while the remaining small parameters are set to zero. As expected, a gap opens at the Dirac points.

It is instructive to consider the gap opening in the FM phase from the point-of-view of the $k\cdot p$ Hamiltonian obtained by linearizing around K and K' ($\hbar=1$):
\begin{equation}
H({\bs k})=v\left(k_x\sigma_x\tau_z+k_y\sigma_y\right)+m_{\epsilon}\sigma_z
\label{eq:HDiracBO}
\end{equation}
Here, $\vec{\sigma}$ are the Pauli matrices acting on the pseudo-spin degree of freedom and $\vec{\tau}$ are the Pauli matrices acting on the valley degree of freedom. (We ignore the physical spin for the fully spin polarized system.) The Fermi velocity is given by $v=3ta/4$ where $a=\sqrt{2/3}a_0$ is the bond length of the projected honeycomb lattice. Importantly, the bond-order wave introduces a mass parameter $m_{\epsilon}=3\epsilon/2$ which has the same sign in the two valleys. It therefore acts similarly to a staggered sublattice potential.

\section{Hartree-Fock Calculations for $U\neq0$.}
\label{sec:HF}
\subsection{Multi-orbital Hubbard model}
As we discussed earlier in this work, the fact that the fully relaxed band structure is qualitatively very similar to the the unrelaxed one presented earlier implies that the Hartree-Fock predictions of Ref.~[\onlinecite{Ruegg:prb12}] are not expected to be qualitatively changed. Indeed, we have verified there is negligible {\em quantitative} change to the phase diagram reported in Fig.~7 of Ref.~[\onlinecite{Ruegg:prb12}]. 

In this section, we instead explore the impact of an external strain along [001] on the various symmetry-broken phases obtained in the mean-field treatment of the interacting system. Our Hartree-Fock calculations follow Ref.~[\onlinecite{Ruegg:prb12}] in which an on-site interaction\cite{Mizokawa:prb96,Imada:rmp98}
\begin{eqnarray}
H_{\rm int}&&=\sum_{\bs r}\Big[U\sum_ {\alpha}n_{{\bs r}\alpha\uparrow}n_{{\bs r}\alpha\downarrow}+(U'-J)\sum_{\alpha>\beta,s}n_{{\bs r}\alpha s}n_{{\bs r}\beta s}\nonumber\\
&&+U'\sum_{\alpha\neq \beta}n_{{\bs r}\alpha\uparrow}n_{{\bs r}\beta\downarrow}+J\sum_{\alpha\neq \beta}d_{{\bs r}\alpha\uparrow}^{\dag}d_{{\bs r}\beta\uparrow}d_{{\bs r}\beta\downarrow}^{\dag}d_{{\bs r}\alpha\downarrow}\nonumber\\
&&+I\sum_{\alpha\neq \beta}d_{{\bs r}\alpha\uparrow}^{\dag}d_{{\bs r}\beta\uparrow}d_{{\bs r}\alpha\downarrow}^{\dag}d_{{\bs r}\beta\downarrow}\Big],
\label{eq:Hint}
\end{eqnarray}
is used.  We assume the following relations between the Slater-Kanamori interaction parameters: $U'=U-2J$ and $I=J$. These are valid in free space and believed to be approximately true in the solid state environment. The total multi-orbital Hubbard Hamiltonian for the $e_g$ electrons is given by
\begin{equation}
H=H_0+H_z+H_{\rm int},
\label{eq:Htot}
\end{equation}
where $H_0$ is the tight-binding Hamiltonian given in Eq.~\eqref{eq:H0} and the effect of the strain is included by the orbital-dependent local energy splitting $H_z$ [Eq.~\eqref{eq:Hz}]. For the tight-binding model $H_0$, we keep the two largest parameters $t$ and $t'$ and set the remaining small parameters to zero.

\subsection{Connection to previous results}
\begin{figure}
\includegraphics[width=1\linewidth]{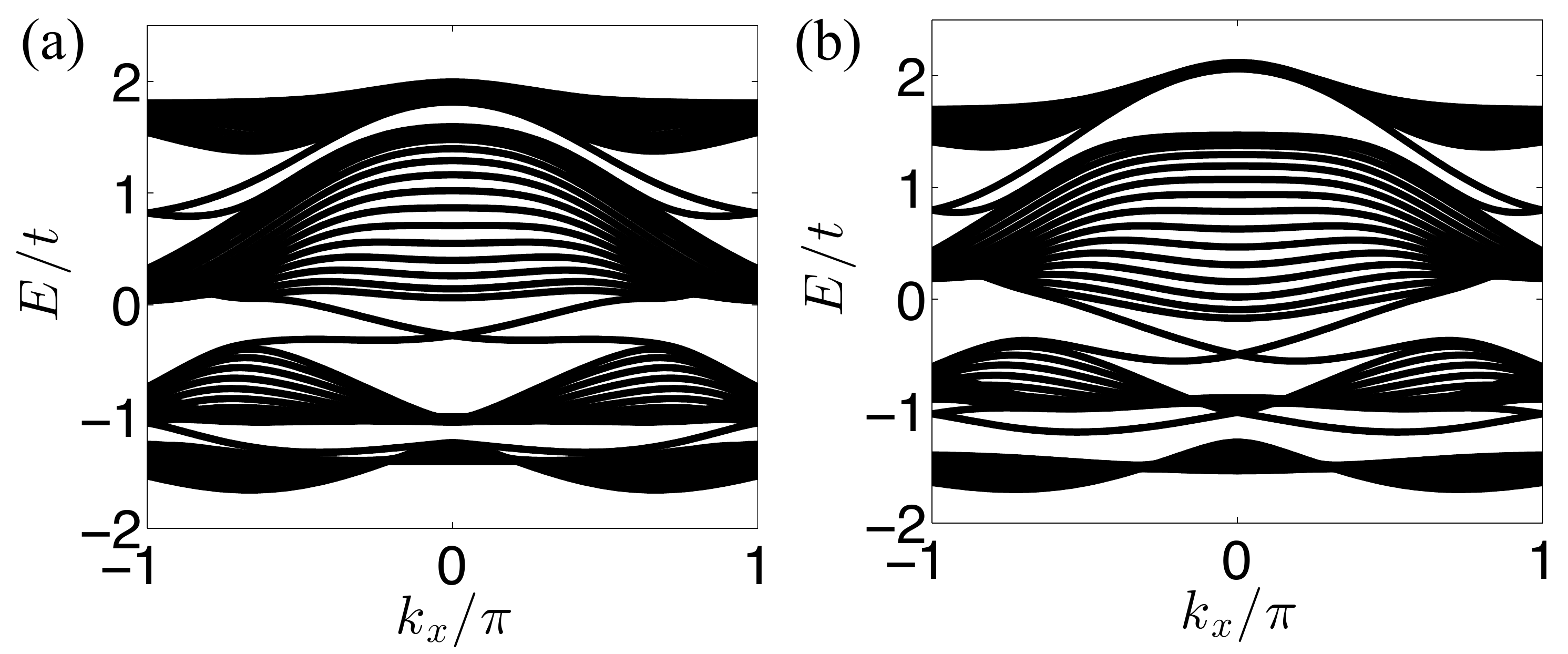}
\caption{Edge states in the QAH model $H_{QAH}$ for a strip with zig-zag edges. (a) External strain $\alpha_z=0$. (b) External strain $\alpha_z=0.25t$. Parameters are $t'=0.1t$, $\tilde{\chi}=0.2t$ and the width of the strip contains $L=31$ sites.}
\label{fig:edges}
\end{figure}

The interacting Hamiltonian Eq.~\eqref{eq:Htot} for the unstrained lattice ($\alpha_z=0$) has been studied previously within the Hartree-Fock approximation and the phase diagram has been worked out for various combinations of interaction parameters.\cite{Yang:prb11a,Ruegg11_2,Ruegg:prb12} A particularly interesting result for intermediate to strong interactions is the observation of a spontaneously generated quantum anomalous Hall (QAH) phase which is accompanied by ordering in complex orbitals within a ferromagnetic (FM) phase. As the FM phase (which appears for larger $J/U$ values) is fully spin polarized, the Fermi energy is placed right at the Dirac points. Physically, the QAH phase then appears as a result of a gap opening at these Dirac points with opposite sign of the mass parameter in the two valleys near K and K$'$.\cite{Haldane:prl88} This gap opening is induced by spontaneous ordering of complex orbitals which is signaled by a non-vanishing expectation value of the $y$-component of the pseudo-spin-1/2 associated with the orbital degree of freedom\cite{Ruegg11_2}
\begin{equation}
\chi=\sum_{s}\langle {\vec{d}_{s,{\bs r}}}^{\dag}\sigma_y\vec{d}_{s,{\bs r}}\rangle=\langle {\vec{d}_{\uparrow,{\bs r}}}^{\dag}\sigma_y\vec{d}_{\uparrow,{\bs r}}\rangle\neq 0,
\label{eq:chi}
\end{equation}
where $\sigma_y$ is the second Pauli matrix and for the last equation we assumed $n_{\uparrow}=1$. In the presence of such an order parameter, the mean-field Hamiltonian acquires a term \cite{Ruegg11_2}
\begin{equation}
H_{\chi}=-\tilde{\chi}\sum_{{\bs r},s}\vec{d}_{s,{\bs r}}^{\dag}\sigma_y\vec{d}_{s,{\bs r}}.
\label{eq:Hchi}
\end{equation}
where the orbital field is determined self-consistently via Eq.~\eqref{eq:chi} and
\begin{equation}
\tilde{\chi}=\frac{\chi}{4}(U-3J).
\end{equation}
The on-site term $H_{\chi}$ opens a gap at the Dirac points. In the $k\cdot p$ Hamiltonian, it enters as a mass parameter $m_{\chi}$ with opposite sign in the two valleys:
\begin{equation}
H({\bs k})=v\left(k_x\sigma_x\tau_z+k_y\sigma_y\right)+m_{\chi}\sigma_z\tau_z,
\label{eq:HDirac_chi}
\end{equation}
where $m_{\chi}=\tilde{\chi}$.
The resulting mean-field band structure is topologically non-trivial displaying a spontaneous quantum anomalous Hall effect with Hall conductivity $\sigma_{xy}=e^2 \nu/h$ where $\nu$ is the Chern number
\begin{equation}
\nu=\frac{1}{2\pi}\int_{\rm BZ}\!\!d^2k\ \Omega({\bs k})=\pm 1.
\end{equation}
$\Omega({\bs k})$ denotes the Berry curvature which is obtained from the mean-field Bloch functions as\cite{Xiao:rmp10}
\begin{equation}
\Omega(k_x,k_y)=i\sum_{n\ \rm{ occ.}}\epsilon_{lm}\langle \partial_{k_l} u_n({\bs k})|\partial_{k_m} u_n({\bs k})\rangle,
\end{equation}
$\epsilon_{lm}$ is the fully anti-symmetric tensor and the sum runs over the occupied bands. The non-trivial Chern number $\nu=\pm 1$ implies the existence of a chiral edge state. Figure \ref{fig:edges} shows the spectrum obtained by studying the model $H_{QAH}=H_0+H_{\chi}+H_z$ on a strip with two zig-zag edges for zero external strain and for $\alpha_z=0.25t$. As expected, chiral edge states are visible at half filling with and without external strain while at quarter filling, the external strain drives a transition to a trivial insulator.
 
These results bear a similarity with Haldane's honeycomb lattice model\cite{Haldane:prl88} which also realizes a non-trivial Chern number $\nu=\pm 1$ in the absence of an external magnetic field. However, the Haldane model is a single-orbital model for spinless fermions and the non-trivial Berry phases appear as a consequence of a complex second-neighbor hopping amplitude.\cite{Haldane:prl88} On the other hand, $H_{QAH}$ involves two orbitals and the non-trivial Berry phases in the QAH phase appear due to ordering in complex orbitals which is apparent in the local term Eq.~\eqref{eq:Hchi}. As discussed in Refs.~\onlinecite{Wu:2008,ZhangM:2011} for the closely related planar $p$-orbital model, the QAH phase is realized at half-filling if $\tilde{\chi}<3t/2$. For $\tilde{\chi}=3t/2$, the gap at half filling closes at $\Gamma$ and a trivial insulator appears for $\tilde{\chi}>3t/2$. In this limit, $H_{QAH}$ essentially describes two copies of the Haldane model separated by $\tilde{\chi}$ and the complex second-neighbor hopping appears in second order perturbation theory in $t/\tilde{\chi}$.\cite{Wu:2008}

Finally, we note that in the vicinity of the QAH/FM phase, Hartree-Fock also predicts\cite{Ruegg:prb12} a gapless FM phase and a gapped $\sqrt{3}\times\sqrt{3}$ AFO/FM phase where a coupling between the two Dirac cones at K and K' is generated by orbital order which triples the unit cell. Furthermore, reducing the value of $J/t$ leads to an antiferromagnet with ferro-orbital order (FO/AFM).

\subsection{Effect of external strain}
\begin{figure}[htb]
\includegraphics[width=0.8\linewidth]{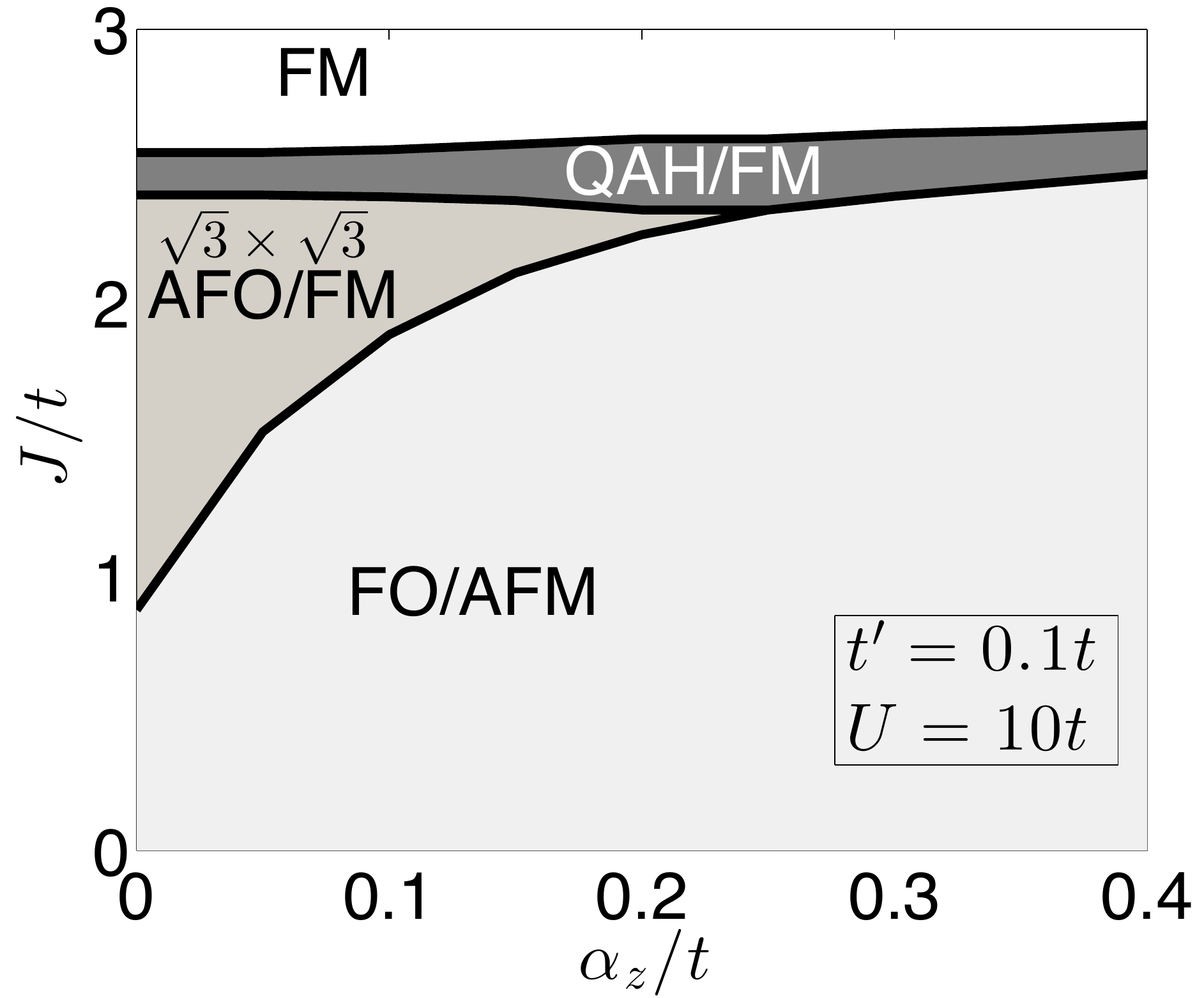}
\caption{Phase diagram as a function of orbital strain $\alpha_z$ [see Eq.~\eqref{eq:Hz}] and the Hund's rule coupling $J/t$ for fixed $U=10t$.  For $x=0.05$, $\alpha_z\approx 0.15$eV $\approx t/4$.  Compared to the unstrained (and numerically similar fully relaxed) case, the AFM is more favored at the expense of the $\sqrt{3} \times \sqrt{3}$ AFO/FM phase.  The parameter regime of the topological quantum anomalous Hall state, QAH/FM, is essentially unaffected. }
\label{fig:strained_phase_diagram}
\end{figure}

In the following, we generalize the previous Hartree-Fock studies to include the effect of external [001] strain. We obtain the phase diagram shown in Fig.~\ref{fig:strained_phase_diagram}.  Based on previous work of the authors, it is known that the phases have a very weak dependence on the value of $t'/t$, as this ratio is small.\cite{Ruegg:prb12} We therefore fix $t'/t=0.1$ and study the phase diagram as a function of the stain-induced $e_g$ orbital splitting, $\alpha_z$, and the Hund's coupling $J$, for fixed $U=10t$, which is a reasonable estimate for LaNiO$_3$.\cite{Ruegg:prb12} Our main result is that the splitting of the $e_g$ orbitals tends to suppress the AFO/FM phase in favor of the FO/AFM phase. Physically, this is because the $e_g$ splitting favors ferro-orbital order, which then biases the system in favor of AFM spin interactions. Our predictions for the topological QAH/FM state with Chern number $\nu=\pm1$ remain {\em quantitatively} similar to the unrelaxed and fully relaxed cases because of the robustness of the Dirac point under the [001] strain, as illustrated by Eq.~\eqref{eq:HDirac_strain}. 

It is interesting to note that recent experiments on the LaNiO$_3$ bilayer grown along [111] did not report robust FM spin order.\cite{Middey:apl12} There is presently no crystal structure data available on this system so it is unclear if symmetry-breaking strain is playing an important role in the physics, but at least the experimental results are roughly consistent with the Hartree-Fock calculations for both the fully relaxed and [001] strained system which identify the most likely phases for the system as the $\sqrt 3 \times \sqrt 3$ AFO/FM and FO/AFM because physical parameters place the system near this phase boundary.  If $J/U,J/t$ were roughly 50-70\% larger the QAH/FM would be a likely candidate as well.  However, it is unclear how one might ``tune" the experimental system to achieve this regime. Including fluctuation effects beyond the Hartree-Fock mean-field approximation typically moves critical interaction strengths to smaller values. This would have the effect of pushing the phase boundaries in Fig.~\ref{fig:strained_phase_diagram} to larger $J/t$ values (since $U=10t$ is fixed) which would tend to favor the FO/AFM phase and disfavor the QAH/FM phase. This expectation is also consistent with the most current experimental results that do not report FM order.

\subsection{Effect of a breathing distortion}
Finally, we mention that a possible breathing distortion as discussed in Secs.~\ref{sec:LSDAU} and \ref{sec:BO} does likely affect the potential QAH/FM phase. This is apparent from the effective models Eqs.~\eqref{eq:HDiracBO} and \eqref{eq:HDirac_chi} where it is seen that the breathing distortion competes with the QAH/FM phase. Indeed, the gap closes if $m_{\epsilon}=m_{\chi}$ which indicates a transition from the topological to the trivial phase triggered by a structural transition. However, because the breathing distortion only appears within LSDA+$U$, we expect a complex interplay between interaction effects and structural distortions which is beyond the scope of the present paper. We leave this interesting problem for a future study.

\section{Experimental identification of topological states}
\label{sec:exp}
From the point-of-view of experimentally identifying topological states of matter in oxide heterostructures, it is important to emphasize that the most natural experiments for identifying two-dimensional topological states with one-dimensional edge modes involve transport.\cite{Konig:sci07,Roth:sci09,Chang:sci13}  The two dimensional topological systems we discussed in this paper are gapped in the bulk, but possess gapless one-dimensional boundary excitations that dominate the low-energy response of the system.  A number of theories based on interacting one-dimensional models have made predictions for a wide range of transport scenarios,\cite{Maciejko:prl09,Xu:prb06,Strom:prl09,Zyuzin:prb10,Fiete:prb06,Wu:prl06,Fiete_transport:prb05,Huo:prl09} and many of them should be applicable to the edge of two-dimensional topological states formed at oxide interfaces.  The key experimental challenge may be ``patterning" and ``contacting" the sample in a way convenient to perform the most desirable transport experiments.  

It appears to us that angle resolve photoemission spectroscopy would be extremely challenging on these samples as the one-dimensional edge signal would likely be rather weak and difficult to detect.  Even for the transport measurements, a promising experimental strategy might be to form a superlattice with bilayers sufficiently widely separated that they are uncoupled from each other.  In this case, transport measurements would pick up a signal of $N$ bilayers ``in parallel".  For known $N$, one could then verify that the conductance scales as expected with $N$: $2 N e^2/h$ for QSH and $N e^2/h$ for QAH with Chern number one for each bilayer.

\section{Conclusions}
\label{sec:conclusions}

In this work we have extended earlier theoretical studies\cite{Yang:prb11a,Ruegg11_2,Ruegg:prb12} on the LaNiO$_3$ [111] bilayer system to include the effects of lattice relaxation and strain on the predicted phase diagrams.  By constraining the in-plane lattice constants to their bulk value and allowing for out-of-plane stretching along with rotations of the oxygen octahedral cages we have found that the fully relaxed band structure obtained within the LDA/GGA approximation to DFT is very similar (only 10-15\% change in tight-binding fit parameters) to the ideal cubic structure. As a result, the previously predicted phase diagrams that used this band structure as an input to a Hartree-Fock calculation are left essentially unchanged, even {\em quantitatively}. This is true both at weak interactions where the quadratic band-touching at the $\Gamma$ point is perturbatively unstable to the spontaneous opening of a gap and the formation of topological states,\cite{Ruegg11_2} and at stronger coupling where a fully polarized ferromagnetic state opens a gap at Dirac points near K and K$'$ to transition to a QAH state.\cite{Ruegg:prb12} Based on experimentally known parameters for LaNiO$_3$, the latter possibility (gapping a Dirac point in a fully spin-polarized state to obtain the QAH state) is likely the most relevant candidate topological transition in the [111] bilayer. 

In addition to the essentially unchanged phase diagrams in the fully relaxed case, we also computed the oxygen layer separation in the (LaNiO$_3$)$_2$/(LaAlO$_3$)$_{10}$ heterostructure and found a compression near the interface and an expansion in the LaNiO$_3$ bilayer. We also computed the layer-resolve oxygen tilt angle in the heterostructure.  These results could be useful in future experimental efforts to ``design" [111]-grown materials with particular octahederal tilts.  

Finally, we found two types of lattice distortions which do compete with topological phases: (i) a symmetry-breaking strain applied along the [001] cubic axis and (ii) a breathing distortion of the oxygen octahedra. A symmetry breaking strain opens a gap at the quadratic band touching at the $\Gamma$ point and hence destroys a topological phase appearing as a weak coupling instability.\cite{Ruegg11_2} However, for the studied range of external strain, the Dirac points, which are relevant to the fully polarized system, remain intact. As a result, instabilities of the gapless fully polarized system to a gapped topological phase persists.\cite{Ruegg:prb12} We explicitly confirmed this expectation by computing a new phase diagram for the [001] strained system over the parameter regime most relevant to LaNiO$_3$ using the Hartree-Fock approximation to an effective Hubbard model which includes the effect of strain. We indeed found that the region of the topological QAH/FM phase is nearly unchanged compared to the fully relaxed and unrelaxed system (a reflection of the stability of the Dirac point).\cite{Ruegg:prb12} However, the FO/AFM phase tends to out-compete the $\sqrt 3 \times \sqrt 3$ AFO/FM as the [001] strain is increased. 
As opposed to the symmetry-breaking strain, we found that a possible breathing distortion, as predicted in the LSDA+$U$ calculation, would compete with a topological phase in the fully polarized system. But because the most recent experimental results\cite{Middey:apl12} do not report inequivalent Ni sites, the experimental relevance of this observation is unclear.

We hope our results which discusses the robustness to the predicted phase diagrams will further spur experimental efforts to search for novel, including topological, phases in the [111]-grown transition metal oxide systems.

\acknowledgements
We thank Mehdi Kargarian and Xiang Hu for helpful discussions. 
We gratefully acknowledge financial support through ARO Grant No.~W911NF-09-1-0527, NSF Grant No.~DMR-0955778, and though grant W911NF-12-1-0573 from the Army Research Office with funding from the DARPA OLE Program. C.M.~and A.A.D.~acknowledge the support provided through Scientific Discovery through Advanced Computing (SciDAC) program funded by U.S. Department of Energy, Office of Science, Advanced Scientific Computing Research and Basic Energy Sciences under award number DESC0008877. A.R.~acknowledges support through the Swiss National Science Foundation. The authors acknowledge the Texas Advanced Computing Center (TACC) at The University of Texas at Austin for providing the necessary computing resources. URL: \url{http://www.tacc.utexas.edu}.

\bibliography{pyro111}

\end{document}